\documentclass[12pt]{article}

\usepackage{amssymb,amsmath,amsbsy,amsthm,cite}
\usepackage{latexsym,graphicx}
\usepackage{authblk}

\usepackage{hyperref}

\newcommand{\cE}{\mathcal{E}}
\newcommand{\cF}{\mathcal{F}}

\newcommand{\cI}{\mathcal{I}}

\newcommand{\cO}{\mathcal{O}}

\newcommand{\cQ}{\mathcal{Q}}

\newcommand{\cS}{\mathcal{S}}

\newcommand{\cU}{\mathcal{U}}

\newcommand{\cW}{\mathcal{W}}

\newcommand{\ee}{\,{\rm e}}
\newcommand{\ii}{{\rm i}}
\newcommand{\vx}{{\bf x}}
\newcommand{\vy}{{\bf y}}
\newcommand{\vu}{{\bf u}}
\newcommand{\vn}{{\bf n}}

\newcommand{\vnu}{\boldsymbol{\nu}}
\newcommand{\vQ}{{\bf K}}
\newcommand{\vk}{{\bf k}}
\newcommand{\ve}{{\bf e}}

\newcommand{\vp}{{\bf p}}
\newcommand{\vzero}{{\bf 0}}

\newcommand{\N}{{\mathbb N}}
\newcommand{\Z}{{\mathbb Z}}
\newcommand{\R}{{\mathbb R}}

\newcommand{\pdag}{^{\phantom\dag}}
\newcommand{\define}{\stackrel{\mbox{{\tiny def}}}{=}}

\newcommand{\tintUnder}[1]{\;\,\tilde{\!\!\!\int\limits_{{#1} }}}

\newcommand{\xxa}{\stackrel {\scriptscriptstyle \times}{\scriptscriptstyle \times} \!}
\newcommand{\xxe}{\! \stackrel {\scriptscriptstyle \times}{\scriptscriptstyle \times}}

\newcommand{\ud}{\mathrm{d}}

\newcommand{\tPiL}{\Bigl(\frac{2\pi}{L}\Bigr)}

\newcommand{\LtPi}{\Bigl(\frac{L}{2\pi}\Bigr)} 

\newcommand{\ta}{\tilde{a}}
\newcommand{\BZ}{{BZ}}

\newcommand{\Hubb}{H_{{\text{Hubb}}}\pdag}
\newcommand{\HubbKin}{H_{{\text{Hubb}}}^{(0)}}
\newcommand{\HubbInt}{H_{{\text{Hubb}}}^{(1)}}

\newcommand{\Pair}{P}
\newcommand{\Charge}{C}
\newcommand{\SpinZ}{{S}}
\newcommand{\Spin}{{\bf{S}}}
\newcommand{\ZM}{Q}
\newcommand{\HEightFlavor}{H_{8{\text{-}}f}}
\newcommand{\HPartContLim}{H_{eff}}
\newcommand{\HWZW}{H}
\newcommand{\HWZWKin}{H_0}
\newcommand{\HWZWInt}{H_1}
\newcommand{\HWZWantnod}{H_{na}^{'}}
\newcommand{\HMattis}{H_M}
\newcommand{\Hn}{H_{n}}
\newcommand{\Ha}{H_{a}}
\newcommand{\Hantnod}{H_{na}}
\newcommand{\TruncHubb}{\tilde H_{{\text{Hubb}}}\pdag}
\newcommand{\TruncHubbKin}{\tilde H_{{\text{Hubb}}}^{(0)}}
\newcommand{\TruncHubbInt}{\tilde H_{{\text{Hubb}}}^{(1)}}

\title{\Large{\bf{Partial continuum limit of the 2D Hubbard model}}}

\date{\vspace{-1.0cm}\small August 14, 2012\vspace{0.2cm}}

\author[1,*]{Jonas de Woul}
\author[1,\dag]{Edwin Langmann}
\affil[1]{Department of Theoretical Physics, Royal Institute of Technology KTH\newline SE-106 91 Stockholm, Sweden \vspace{2mm}}

\begin{document}

\maketitle

\theoremstyle{plain}
\newtheorem{theorem}{Theorem}[section]
\newtheorem{proposition}[theorem]{Proposition}
\newtheorem{lemma}[theorem]{Lemma}
\newtheorem{corollary}[theorem]{Corollary}
\newtheorem{result}[theorem]{Result}
\newtheorem{conjecture}[theorem]{Conjecture}
\theoremstyle{definition}
\newtheorem{definition}[theorem]{Definition}
\newtheorem{approximation}[theorem]{Approximation}
\theoremstyle{remark}
\newtheorem{remark}[theorem]{Remark}

\let\oldthefootnote\thefootnote
\renewcommand{\thefootnote}{\fnsymbol{footnote}}
\footnotetext[1]{Electronic address: {\tt jodw02@kth.se}}
\footnotetext[2]{Electronic address: {\tt langmann@kth.se}}
\let\thefootnote\oldthefootnote

\vspace{-1.5cm}

\begin{abstract}
An effective quantum field theory of the 2D Hubbard model on a square lattice near half-filling is presented and studied. This effective model describes so-called nodal- and antinodal fermions, and
it is  derived from the lattice model using a certain partial continuum limit. It is shown that the nodal fermions can be bosonized, which leads to spin-charge separation and a 2D analogue of a Wess-Zumino-Witten model. A bosonization formula for the nodal fermion field operator is obtained, and an exactly solvable model of interacting 2D fermions is identified. Different ways of treating the antinodal fermions are also proposed. 
\end{abstract}

\noindent \noindent {\bf Remark added on September 20, 2023:} {\em This paper was included in the PhD thesis of the first author, who defended his PhD on December 16, 2011 (this thesis: ``Fermions in two dimensions and exactly solvable models,'' is available on \href{http://kth.diva-portal.org/}{http://kth.diva-portal.org/}). We planned to publish this paper, but for some reason or another this did not happen then. 

We make this paper available on the arXiv in the form it was on August 14, 2012 (this is a slight update of the version that appeared in the above-mentioned PhD thesis). }

\section{\label{Sec:Introduction}Introduction}

Advancing our computational understanding of the Hubbard model \cite{Hubbard:1963,Kanamori:1963,Gutzwiller:1963} is an important but challenging problem in the theory of many-electron systems. As one of \emph{the} minimal models for strongly correlated electrons, its ground state is believed to describe various charge-ordered-, magnetic- and superconducting phases for different parameter values and spatial dimensionality \cite{Lieb:1995,Scalapino:2007}. 
The Hamiltonian can be represented as
\begin{equation}
\label{Hubbard Hamiltonian short-hand notation}
\Hubb = -\sum\limits_{\alpha=\uparrow,\downarrow}\sum\limits_{i,j}t_{ij}\pdag c_{i,\alpha}^\dag c_{j,\alpha}\pdag+ U\sum\limits_{i}n_{i,\uparrow}n_{i,\downarrow}  
\end{equation} 
with operators $c_{i,\alpha}^\dag$ and $c_{i,\alpha}\pdag$ describing the creation- and annihilation of a fermion with spin projection $\alpha$ at lattice site $i$, $n_{i,\alpha} = c_{i,\sigma}^\dag c_{i,\sigma}\pdag$ the corresponding density operators, $U\geq 0$ the strength of the screened Coulomb repulsion, and $t_{ij}$ the hopping matrix elements.
Of particular interest for the high-Tc problem of the cuprate superconductors \cite{Anderson:1987,Dagotto:1994,LeeNagaosaWen:2006} is the two-dimensional (2D) model on a square lattice, which is the focus of the present paper. At half-filling and sufficiently large $U$, there is by now compelling evidence that the model is a Mott insulator \cite{ImadaFujimoriTokura:1998} with strong antiferromagnetic correlations, as seen for example in rigorous Hartree-Fock- \cite{BachLiebSolovej:1994,BachPoelchau:1996} and quantum Monte Carlo studies \cite{HirschTang:1989,Varney:2009}. Less is known away from half-filling. Numerical Hartree-Fock studies find a plethora of inhomogeneous solutions like polarons, different types of domain walls or stripes, vortex-like structures and ferromagnetic domains; see \cite{Verges:1991} and references therein. 
Furthermore, renormalization group studies at weak coupling show Fermi-liquid behavior far from half-filling \cite{BenfattoGiulianiMastropietro:2006}, and strong tendencies towards antiferromagnetism and $d$-wave superconductivity close to half-filling \cite{ZanchiSchulz:2000,HalbothMetzner2:2000,HonerkampSalmhofer:2001,HonerkampSalmhoferFurukawaRice:2001,Metzner:2011}; 
similar results are obtained from quantum cluster methods \cite{SenechalLavertuMaroisTremblay:2005,Maier:2005}. Still, few definitive conclusions can be drawn for arbitrary coupling strength.

This level of uncertainty may be contrasted with the corresponding situation in one dimension. The 1D Hubbard model with nearest-neighbor hopping is integrable and can be solved exactly using Bethe ansatz; see \cite{Essler:2005} and references therein. More general 1D lattice models of fermions can be successfully studied using numerical methods, e.g.\ the density matrix renormalization group \cite{White:1992}. 
An alternative approach is to perform a particular continuum limit away from half-filling that leads to a simplified model that can be studied by analytical methods. This limit involves linearising the tight-binding band relation at the non-interacting Fermi surface points and ``filling up the infinite Dirac sea of negative energy states''. For spinless fermions one obtains the (Tomonaga-)Luttinger model \cite{Tomonaga:1950,Luttinger:1963}, which can be solved using bosonization \cite{MattisLieb:1965}; in particular, all thermodynamic Green's functions can be computed  \cite{Theumann:1967,Dover:1968,DzyaloshinskiiLarkin:1974,EvertsSchulz:1974,LutherPeschel:1974,Fogedby:1976,HeidenreichSeilerUhlenbrock:1980}. 
Generalizing to arbitrary interacting fermion models away from half-filling leads to the notion of the Luttinger liquid \cite{Haldane:1981} -- the universality class of gapless Fermi systems in one dimension (see e.g.\ \cite{Voit:1995} for review). Furthermore, spinfull systems like the 1D Hubbard model can be studied using both abelian- and non-abelian bosonization, with the latter leading to a Wess--Zumino--Witten-type (WZW) model \cite{Witten:1984,Affleck:1990}. We note that bosonization has a rigorous mathematical foundation, see e.g.\ \cite{CareyRuijsenaars:1987,vonDelftSchoeller:1998}.

The idea of applying bosonization methods in dimensions higher than one goes back to pioneering work of Luther \cite{Luther:1979}, and was popularized by Anderson's suggestion that the Hubbard model on a square lattice might have Luttinger-liquid behavior away from half-filling \cite{Anderson:1990}. Consider for example a gapless system with a square Fermi surface. Let $k_\parallel$ and $k_\perp$ denote fermion momenta parallel and perpendicular, respectively, to a face of the square. Following \cite{Luther:1979}, one would treat $k_\parallel$ as a flavor index, extend $k_\perp$ to be unbounded, and fill up the Dirac sea such that all states $k_\perp<0$ are filled. The system can then be bosonized by the same methods used in one dimension. Unfortunately, in this approach only density operators with momentum exchange in the perpendicular direction behave as bosons, while operators with exchange in the parallel direction do not have simple commutation relations. Yet, Mattis \cite{Mattis:1987} proposed a 2D model of spinless fermions with density-density interactions, containing momentum exchange in all directions, that he claimed was solvable using bosonization. The Hamiltonian of Mattis' model had a kinetic energy term with a linear tight-binding band relation on each face of a square Fermi surface, and with a constant Fermi velocity $v_F$ along each face. Mattis rewrote the kinetic energy as a quadratic expression in densities using a generalized Kronig identity, and the Hamiltonian was then diagonalized by a Bogoliubov transformation. 

The exact solubility of Mattis' model can be understood in light of more recent work of Luther \cite{Luther:1994} in which he studied a model of electrons with linear band relations on a square Fermi surface: A notable difference to the 1D case is the huge freedom one has in choosing the accompanying flavor indices when bosonizing. In particular, one may do a Fourier transformation in the $k_\parallel$-direction and then bosonize using a new index flavor $x_\parallel$. In this way, Luther obtained density operators that indeed satisfy 2D boson commutation relations. The price one has to pay for solubility is that $v_F$ needs to be constant on each face, i.e. it cannot depend on $k_\parallel$. The properties of Luther's model were further investigated in \cite{FjaerestadSudboLuther:1999,SyljuaasenLuther:2005}. We also mention Haldane's phenomenological approach to bosonization in higher dimensions \cite{Haldane:1994}, which has been further pursued by various groups \cite{HoughtonMarston:1993,CastroNetoFradkin2:1994,Hlubina:1994}, 
and functional integral approaches to bosonization \cite{FrohlichGotschmannMarchetti:1995,KopietzSchonhammer:1996}; none of these will be followed here.

Returning to the 2D Hubbard model, consider momentarily the half-filled square lattice with nearest-neighbor (nn) hopping only. The tight-binding band relation relevant in this case is\footnote{We write $\vk=(k_1,k_2)$ for fermion momenta, $t>0$ is the nn hopping constant, and we set the lattice constant $a=1$ in this section.} $\epsilon(\vk)=-2t[\cos(k_1)+\cos(k_2)]$, which gives a square (non-interacting) Fermi surface at half-filling. The functional form of $\epsilon(\vk)$ varies significantly over this surface: In the so-called {\em nodal} regions of the Brillouin zone near the midpoints $(\pm\pi/2,\pi/2)$ and $(\pi/2,\pm\pi/2)$ of the four faces, the band relation is well represented by a linear approximation in the perpendicular direction to each face. In contrast, at the corner points $(\pm\pi,0)$ and $(0,\pm\pi)$ in the so-called {\em antinodal} regions, $\epsilon(\vk)$ has saddle points. This makes taking a constant Fermi velocity along each face a questionable approximation. Furthermore, we know that the van-Hove singularities associated with these saddle points, and the nesting of the Fermi surface, give various ordering instabilities that can lead to gaps \cite{Schulz:1989}. Of course, going away from half-filling or including further neighbor hopping can bend the Fermi surface away from these points. Moreover, even if the concept of a Fermi surface survives at intermediate- to strong coupling, the interaction is likely to renormalize the surface geometry \cite{HalbothMetzner:1997}. Nonetheless, the fermion degrees of freedom in the nodal- and antinodal regions are likely to play very different roles for the low-energy physics of the Hubbard model. 

In this paper we develop a scheme that improves the bosonization treatments of the 2D Hubbard model mentioned above. The basic idea is to treat nodal- and antinodal degrees of freedom using differing methods. To be specific, we perform a certain {\em partial} continuum limit that only involves the nodal fermions and that makes them amenable to bosonization, while allowing to treat the antinodal fermions by conventional methods like a mean-field- or random phase approximation. This is an extension of our earlier work on the so-called 2D $t$-$t'$-$V$ model of interacting spinless fermions \cite{Langmann1:2010,Langmann2:2010,deWoulLangmann1:2010,deWoulLangmann2:2010}. In the spinless case, the partial continuum limit gives a natural 2D analogue of the Luttinger model consisting of nodal fermions coupled to antinodal fermions \cite{Langmann1:2010,Langmann2:2010}. This effective model is a quantum field theory (QFT) model (by this, we mean that the model has an infinite number of degrees of freedom) and, as such, requires short- and long distance regularizations \cite{Langmann2:2010,deWoulLangmann2:2010}. These regularizations are provided by certain length scale parameters $\ta$ (proportional to the lattice constant) and $L$ (the linear size of the lattice). 
After bosonizing the nodal fermions, one can integrate them out exactly using functional integrals, thus leading to an effective model of antinodal fermions only \cite{Langmann2:2010}. It was shown in \cite{deWoulLangmann1:2010} that this antinodal model allows for a mean field phase corresponding to charge ordering (charge--density--wave), such that the antinodal fermions are gapped and the total filling of the system is near, but not equal to, half-filling. In this {\em partially gapped phase} 
the low-energy properties of the system are governed by the nodal part of the effective Hamiltonian. This nodal model is exactly solvable:\ the Hamiltonian can be diagonalised and all fermion correlation functions can be computed by analytical methods \cite{deWoulLangmann2:2010}. One finds, for example, that the fermion two-point functions have algebraic decay with non-trivial exponents for intermediate length- and time scales. The purpose of this paper is to extend the above analysis to fermions with spin. In the main text we explain the ideas and present our results, emphasizing the differences with the spinless case. Details and technicalities (which are important in applications of our method) are deferred to appendices. One important feature of our method is its flexibility. To emphasize this, the results in the appendices are given for an extended Hubbard model that also includes a nn repulsive interaction.  

In Section~\ref{Sec: The effective field theory}, we summarize our results by giving a formal\footnote{By ``formal'' we mean that details of the short- and long distance regularizations needed to make these models well-defined are ignored; these details are spelled out in other parts of the paper.} description of the effective QFT model that we obtain. 
We then outline how the partial continuum limit is done for the 2D Hubbard model in Section~\ref{Sec:Partial continuum limit}. In Section~\ref{Sect: Bosonization of nodal fermions}, we define the nodal part of the effective model and show how it can be bosonized by operator methods. We also identify an exactly solvable model of interacting fermions in 2D. In Section~\ref{Sect: Integrating out degrees of freedom}, we include the antinodal fermions in the analysis and discuss how different effective actions may be obtained by integrating out either the nodal- or the antinodal fermions. The final section contains a discussion of our results. Computational details, including formulas relating the Hubbard model parameters to the parameters of the effective QFT model, are given in Appendices~\ref{App: Index sets}--\ref{App: Functional integration of nodal bosons}.

{\em Notation}: For any vector $\vu\in\R^2$, we write either $\vu=(u_1,u_2)$ or $\vu= u_+\ve_++u_-\ve_-$, with $u_\pm\define(u_1\pm u_2)/\sqrt{2}$ and $\ve_\pm\define(1,\pm 1)/\sqrt{2}$. We denote the Pauli matrices by $\sigma^i$, $i=1,2,3$, 
the $2\times 2$ unit matrix as $\sigma^0$, and $\sigma^\pm = (\sigma^1 \pm \ii\sigma^2)/2$. Spin quantum numbers are usually written as $\uparrow,\downarrow$, but sometimes also as $\pm$. We write $h.c.$ for the hermitian conjugate. Fermion- and boson normal ordering of an operator $A$ is written $:\! A \!:$ and $\xxa A \xxe$, respectively.
We sometimes use the symbol ``$\define$'' to emphasize that an equation is a definition.
When there is no risk of confusion, we often suppress position- or momentum arguments of operators, writing e.g.\ $\hat J_{r,s}^\dag\hat J_{r',s'}\pdag \define \hat J_{r,s}\pdag(-\vp)\hat J_{r',s'}\pdag(\vp)$.

\section{\label{Sec: The effective field theory}The effective field theory}

Similar to the spinless case \cite{Langmann1:2010,Langmann2:2010,deWoulLangmann1:2010,deWoulLangmann2:2010}, doing the partial continuum limit for the spinfull lattice fermions also leads to a quantum field theory model of coupled nodal- and antinodal fermions, although much richer in structure and complexity. In this section, we outline the different parts of this model while, at the same time, suppress most of the technical details making the model mathematically well-defined. Complete details are given in later sections and in the appendices.

\subsection{A 2D analogue of a Wess--Zumino--Witten model}

As will be shown, the nodal part of the full effective Hamiltonian (see below) has a contribution formally given by (we suppress all UV regularizations in this section)
\begin{equation}
\label{WZW Hamiltonian in position space}
\begin{split}
\HWZW = \int\ud^2 x\,\Bigl(&v_F \sum\limits_{\alpha=\uparrow,\downarrow}\sum\limits_{r,s=\pm}  :\! \psi^\dag_{r,s,\alpha}(\vx)(-\ii r\partial_s)\psi_{r,s,\alpha}\pdag(\vx)\!: + g\bigl(\sum\limits_{r,s=\pm}   J_{r,s}^0J_{r,s}^0 
\\
&+  \sum\limits_{s=\pm}J_{+,s}^0 J_{-,s}^0+\sum\limits_{r,r'=\pm}  J_{r,+}^0J_{r',-}^0 -\sum\limits_{s=\pm} { {\bf J}}_{+,s} \cdot { {\bf J}}_{-,s}-\sum\limits_{r,r'=\pm}  {{\bf J}}_{r,+} \cdot {{\bf J}}_{r',-}
\bigr)\Bigr)
\end{split} 
\end{equation}
with $\partial_{\pm}=\partial/\partial x_\pm$ and $x_\pm$ Cartesian coordinates of $\vx$. The fermion field operators $\psi\pdag_{r,s,\alpha}(\vx)$ obey canonical anticommutator relations $\{\psi\pdag_{r,s,\alpha}(\vx), \psi^{\dag}_{r',s',\alpha'}(\vy)\} = \delta_{r,r'}\delta_{s,s'}\delta_{\alpha,\alpha'}\delta(\vx-\vy)$, etc., and $r,s=\pm$ are certain flavor indices. The coupling constant $g$ is proportional to $U$. 
Furthermore,
\begin{equation}
\label{Current operators in position space}
\begin{split}
J_{r,s}^0(\vx) &= \sum\limits_{\alpha}:\! \psi^\dag_{r,s,\alpha}(\vx)\psi_{r,s,\alpha}\pdag(\vx)\!: 
\\
{\bf J}_{r,s}(\vx) &= \sum\limits_{\alpha,\alpha'}:\!\psi^\dag_{r,s,\alpha}(\vx){\boldsymbol{\sigma}}_{\alpha,\alpha'}\pdag\psi_{r,s,\alpha'}\pdag(\vx)\!:,\qquad \boldsymbol{\sigma}=(\sigma^1,\sigma^2,\sigma^3)
\end{split} 
\end{equation}
are 2D (fermion normal-ordered) density- and (rescaled) spin operators for which the non-trivial commutation relations are given by (again formally)
\begin{equation}
\label{WZW commutation relations in position space}
\begin{split}
\left[J_{r,s}^0(\vx), J_{r,s}^0(\vy)\right]  = & r\frac{1}{\pi\ta\ii}\partial_s \delta \left(\vx - \vy\right)
\\
\left[J_{r,s}^i(\vx), J_{r,s}^j(\vy)\right]  = &  2\ii\sum\limits_{k} {\epsilon_{ijk} J_{r,s}^k (\vx)}\delta\left(\vx - \vy\right)
+ r\frac{1}{\pi\ta\ii}\delta_{i,j}\partial_s\delta\left(\vx - \vy\right)
\end{split}.
\end{equation}
We also set ${\bf S}_{r,s}(\vx)={\bf J}_{r,s}(\vx)/2$. We find by using a particular Sugawara construction that the Hamiltonian in \eqref{WZW Hamiltonian in position space} separates into a sum of independent density- and spin parts (spin-charge separation)
\begin{equation}
\label{Spin-charge separation in position space}
\HWZW = \HWZW_\Charge\pdag + \HWZW_\Spin\pdag
\end{equation}
with
\begin{equation}
\label{Spin and charge parts in position space}
\begin{split}
&\HWZW_\Charge\pdag = \frac{v_F}{2}\int\ud^2 x\, \pi\ta\xxa \Bigl(\sum\limits_{r,s}\bigl(({1 + 2\gamma})J_{r,s}^0J_{r,s}^0 + \gamma J_{r,s}^0 J_{-r,s}^0\bigr) + 2\gamma \sum\limits_{r,r' } {J_{r,+}^0 J_{r',-}^0}\Bigr)\xxe
\\
&\HWZW_\Spin\pdag = \frac{v_F}{2} \int\ud^2 x\, \pi\ta \xxa \Bigl(\sum\limits_{r,s} \bigl({\bf J}_{r,s} \cdot {\bf J}_{r,s}/3  - \gamma {\bf J}_{r,s}\cdot {\bf J}_{-r,s}\bigr)  - 2\gamma \sum\limits_{r,r'} {\bf J}_{r,+} \cdot {\bf J}_{r',-} \Bigr) \xxe
\end{split}
\end{equation}
and with a dimensionsless coupling constant $\gamma\geq0$ proportional to $g$. As is evident from the multiple occurence of the short-distance scale $\ta$ in \eqref{WZW commutation relations in position space} and \eqref{Spin and charge parts in position space}, a proper quantum field theory limit $\ta\to 0^+$ of the effective model can possibly make sense only after certain non-trivial multiplicative renormalizations of observables (and implementing a UV regularization on the Hamiltonian). The algebra in \eqref{WZW commutation relations in position space} and the Sugawara construction leading to \eqref{Spin-charge separation in position space}--\eqref{Spin and charge parts in position space} can naturally be interpreted as giving a WZW-type model in two spatial dimensions.

\subsection{The full nodal--antinodal model}

The full effective Hamiltonian of the nodal-antinodal system is given by
\begin{equation}
\label{Introduction: Effective Hamiltonian near half-filling}
\HPartContLim = \Hn + \Ha + \Hantnod
\end{equation}
with the terms on the right hand side corresponding to a pure nodal part ($n$), a pure antinodal part ($a$), and a nodal-antinodal interaction ($na$), respectively. We find that
\begin{equation}
\label{Nodal Hamiltonian in position space}
\Hn = \HWZW + g_n^\Pair  \int\ud^2 x\,\Pair_{r,s}^\dag(\vx)\cdot\Pair_{r',-s}\pdag(\vx),
\end{equation}
with $H$ defined in \eqref{WZW Hamiltonian in position space}, 
\begin{equation}
\begin{split}
\label{Antinodal Hamiltonian in position space}
\Ha = \int\ud^2 x\,\sum\limits_{r=\pm}\Bigl(&\sum\limits_{\alpha} :\!\psi _{r,0,\alpha }^\dag (\vx)\bigl(rc_F\partial_+\partial_- + c_F'(\partial_+^2+\partial_-^2) - \mu_0\bigr) \psi_{r,0,\alpha}\pdag(\vx)\!: 
\\
&+ g_a^\Charge J_{r,0}\pdag J_{r,0}\pdag+ \tilde g_a^\Charge J_{r,0}\pdag J_{-r,0}\pdag + g_a^\SpinZ {\bf S}_{r,0}\pdag \cdot {\bf S}_{-r,0}\pdag + g_a^\Pair \Pair_{r,0}^\dag  \cdot \Pair_{-r,0}\pdag\Bigr)
\end{split},
\end{equation}
and
\begin{equation}
\label{Nodal-antinodal interaction in position space}
\Hantnod = \int\ud^2 x\, \sum\limits_{r,r',s=\pm} \bigl(g_{na}^\Charge J_{r,s}\pdag J_{r',0}\pdag + g_{na}^\SpinZ {\bf S}_{r,s}\pdag \cdot {\bf S}_{r',0}\pdag + g_{na}^\Pair(\Pair_{r,s}^\dag \cdot \Pair_{r',0}\pdag + h.c.)/2\bigr)
\end{equation}
(the coupling constants are defined in terms of the original Hubbard model parameters in Appendix~\ref{App:Derivation of the effective model}). While the definition of the density- and spin operators for the antinodal fermions in \eqref{Antinodal Hamiltonian in position space} are similar to \eqref{Current operators in position space}, we note that there are no anomalous (Schwinger) terms in their commutation relations (cf. \eqref{WZW commutation relations in position space}). The operators $\Pair_{r,s}^\mu$ in \eqref{Nodal Hamiltonian in position space}--\eqref{Nodal-antinodal interaction in position space} are certain pairing bilinears given by
\begin{equation}
\label{Pairing bilinears in position space}
\begin{split}
&\Pair_{r,s}^0(\vx) =\frac12\sum\limits_{\alpha}\psi\pdag_{r_s,s,\alpha}(\vx)\psi\pdag_{r,s,\alpha}(\vx)
\\
&\Pair_{r,s}^i(\vx) =\frac12\sum\limits_{\alpha,\alpha'}\psi\pdag_{r_s,s,\alpha}(\vx)\sigma_{\alpha,\alpha'}^i\psi\pdag_{r,s,\alpha'}(\vx) \qquad (i=1,2,3) 
\end{split}
\end{equation}
with the flavor index $r_{s}\equiv -r$ and $r_{s}\equiv r$ for nodal- ($s=\pm$) and antinodal ($s=0$) fermions, respectively. 
We note that pairing nodal fermions with opposite flavor (chirality) index $r$ is compatible with pairing momenta $\vk$ with $-\vk$ in the Brillouin zone. The same holds true for antinodal fermions with equal flavor index $r$.

One can use abelian bosonization to rewrite the nodal part of the effective model in terms of boson fields corresponding to charge- and spin degrees of freedom. If one truncates \eqref{WZW Hamiltonian in position space} by only keeping the third spin components in the spin rotation invariant interaction, the remaining part becomes quadratic in these boson fields and can thus be diagonalised by a Bogoliubov transformation. This diagonalisation requires that
\begin{equation}
\label{Constraint on gamma}
0\leq \gamma<1/3
\end{equation}
which translates into constraints on the original Hubbard parameters; one finds that $U/t$ must be bounded from above by a value between ten and twenty. Furthermore, the other spin components and the nodal pairing bilinears in \eqref{Pairing bilinears in position space} can be written in terms of exponentials of the charge- and spin boson fields (cf. bosonization of the 1D Hubbard model; see e.g. \cite{Giamarchi:2004}).

\section{\label{Sec:Partial continuum limit}Partial continuum limit}
Our partial continuum limit of the Hubbard model near half-filling is similar to the one done in \cite{Langmann2:2010} for a lattice model of spinless fermions. In this section, we outline the main steps in this derivation; technical details are given in Appendix~\ref{App:Derivation of the effective model}. 

We consider the two-dimensional Hubbard model with nearest- (nn) and next-nearest neighbor (nnn) hopping on a square lattice with lattice constant $a$ and $(L/a)^2$ lattice sites. The Hamiltonian is defined as (equivalent to \eqref{Hubbard Hamiltonian short-hand notation} up to a chemical potential term)
\begin{equation}
\label{Hubbard Hamiltonian in Fourier space: main}
\Hubb = \sum\limits_{\alpha=\uparrow,\downarrow}\sum\limits_{\vk\in\BZ}\left(\epsilon(\vk)-\mu\right) \hat c_{\alpha}^\dag(\vk) \hat c_{\alpha}\pdag(\vk)
+\frac{U}{2}\left(\frac{a}{L}\right)^2\sum\limits_{\vp} \hat\rho({-\vp})\hat\rho({\vp})
\end{equation}
with the fermion operators normalized such that 
$\{\hat c\pdag_{\alpha}(\vk),\hat c^{\dag}_{\alpha'}(\vk')\} =\delta_{\vk,\vk'}\delta_{\alpha,\alpha'}$, 
\begin{equation}
\label{Tight-binding band relation}
\epsilon(\vk)=-2t\left[\cos \left( k_1a\right) + \cos\left( k_2a\right) \right]-4t'\cos \left( k_1a\right)\cos\left( k_2a\right)
\end{equation} 
the tight-binding band relation, and
\begin{equation}
\label{Density operators: main}
\hat\rho(\vp) =\sum\limits_{\alpha=\uparrow,\downarrow}\sum\limits_{\vk_1\vk_2\in\BZ}\sum_{\vn\in\Z^2} \hat c_{\alpha}^\dag(\vk_1) \hat c_{\alpha}\pdag(\vk_2) \delta_{\vk_1+\vp+2\pi\vn/a,\vk_2}
\end{equation}
Fourier-transformed density operators. We assume that the parameters satisfy the constraints $|t'|\leq t/2$ and $U\geq 0$. The average number of fermions per site, or {\em filling factor}, is denoted by $\nu$. Note that $0\leq \nu \leq 2$, with {\em half-filling} corresponding to $\nu=1$.

\begin{figure}[!htb]
\begin{center}
\includegraphics[angle=0, width=0.3\textwidth, trim= 5cm 4.5cm 5cm 1.5cm]{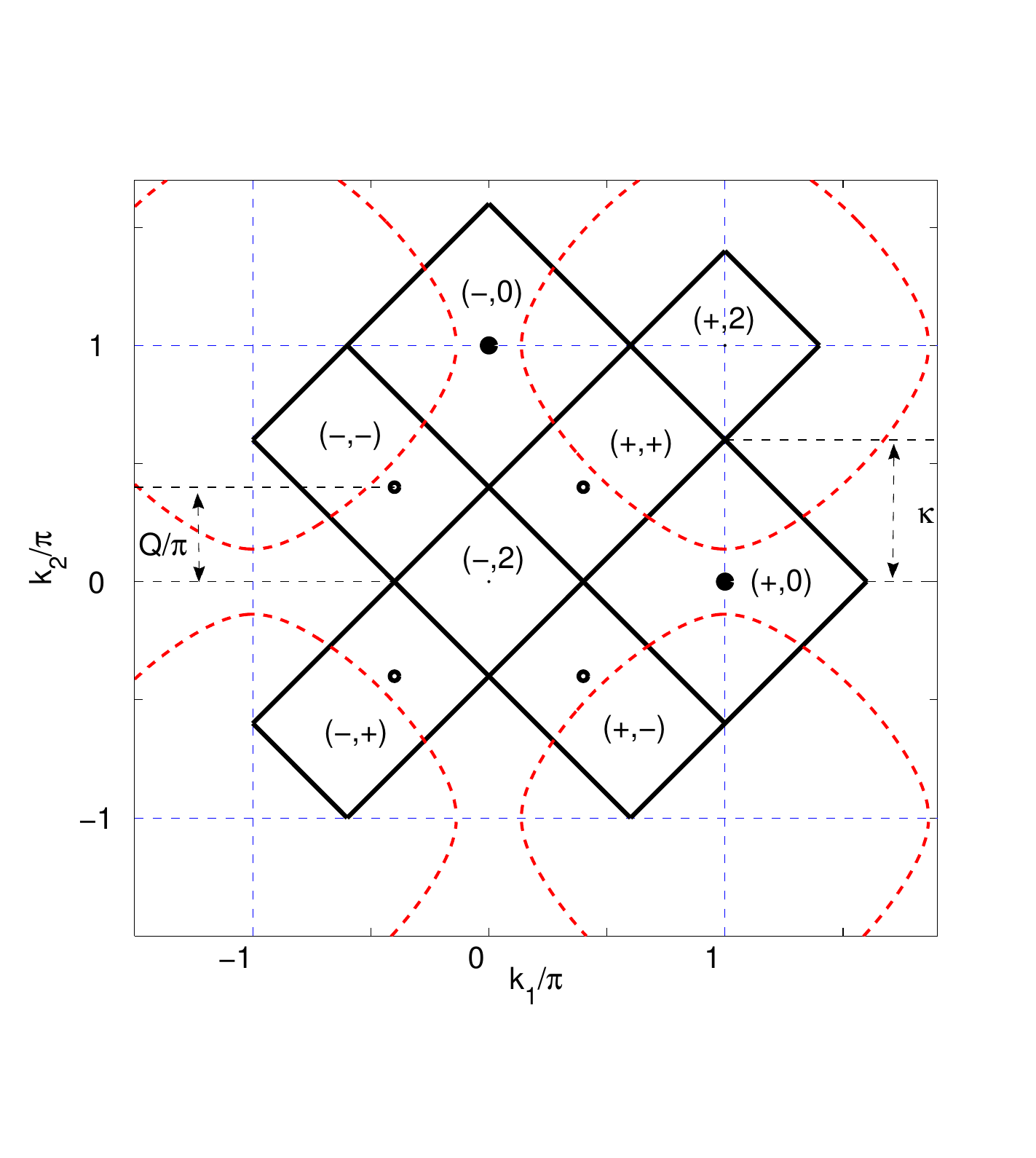} 
\end{center}
\caption{Partition of non-equivalent momenta into eight disjoint regions (rectangles), whose union under suitable translations by reciprocal lattice vectors is the first Brillouin zone. The regions are labelled by pairs of indices $(r,s)$ with $s=0$ corresponding to antinodal fermions, $s=\pm$ nodal fermions, and $s=2$ in- or out fermions. The dashed curves is a superimposed non-interacting Fermi surface corresponding to $t=1$, $t'=-0.2$ and $\mu=-0.51(1)$. We set the lattice constant $a=1$.}
\label{Partition of BZ}
\end{figure}

We choose to classify one-particle degrees of freedom with momenta $\vk$ according to the functional form of $\epsilon(\vk)$ in \eqref{Tight-binding band relation} as discussed in the introduction. This enables us to disentangle fermions that (presumably) play different roles for the low-energy physics of the model. To this end, we introduce eight non-overlapping regions in momentum space identified by pairs of indices $(r,s)$, with $r=\pm$ and $s=0,\pm,2$; see the patchwork of rectangles in Figure~\ref{Partition of BZ}. These regions are defined such that their union is the (first) Brillouin zone, modulo translations of individual momenta by a reciprocal lattice vector. 
We define the eight regions mathematically by associating to each one a fixed point $\vQ_{r,s}$ and a momentum set $\Lambda_{r,s}^*$, such that every momenta in the (first) Brillouin zone can be written uniquely as $\vQ_{r,s} + \vk $ (modulo reciprocal lattice vectors) for some pair of flavor indices $(r,s)$ and momenta $\vk\in\Lambda_{r,s}^*$. The relative size of each region is parameterized by a variable $0\leq\kappa\leq 1$. The precise definitions of the sets $\Lambda_{r,s}^*$ are given in Appendix~\ref{App:Derivation of the effective model} and is further discussed in \cite{Langmann2:2010}. 

The eight regions correspond to three classes of fermion degrees of freedom. We let $s=0$ label so-called antinodal fermions and define $\vQ_{+,0}\define(\pi/a,0)$ and $\vQ_{-,0}\define(0,\pi/a)$. Similarly, we let $s=\pm$ label so-called nodal fermions and define $\vQ_{r,s} = (rQ/a,rsQ/a)$ with a parameter $Q$ close, but not equal, to $\pi/2$. To get a simple geometry, it is useful to also introduce so-called {\em in-} and {\em out} fermions labelled by $s=2$. The corresponding points are $\vQ_{-,2}=(0,0)$ (in) and $\vQ_{+,2}=(\pi/a,\pi/a)$ (out), i.e.\ the center and corners of the Brillouin zone. In the following, one can equally well think of the in- and out fermions as belonging to the nodal fermions. We also define new fermion operators $\hat c_{r,s,\alpha}^{(\dag)}(\vk) = \hat c_{\alpha}^{(\dag)}(\vQ_{r,s} + \vk)$ such that the Hubbard Hamiltonian in \eqref{Hubbard Hamiltonian in Fourier space: main} can be represented as 
\begin{equation}
\Hubb = \HubbKin + \HubbInt
\end{equation}
with
\begin{equation}
\label{Kinetic part in eight-flavor representation: main}
\HubbKin=\sum\limits_{\alpha=\uparrow,\downarrow}\sum\limits_{r=\pm}\sum\limits_{s=0,\pm,2}\sum\limits_{\vk\in\Lambda^*_{r,s}}\left(\epsilon(\vQ_{r,s}+\vk)-\mu+U/2\right)\hat c_{r,s,\alpha}^{\dag}(\vk)\hat c_{r,s,\alpha}\pdag(\vk)
\end{equation}
the free part, and
\begin{equation}
\label{Interaction part in eight-flavor representation: main}
\begin{split}
\HubbInt=U\left(\frac{a}{L}\right)^2\sum\limits_{r_j,s_j}\sum\limits_{\vk_j\in\Lambda^*_{r_j,s_j}}\sum_{\vn\in\Z^2}\delta_{\vQ_{r_1,s_1}-\vQ_{r_2,s_2}+\vQ_{r_3,s_3}-\vQ_{r_4,s_4}+\vk_1-\vk_2+\vk_3-\vk_4,2\pi\vn/a} 
\\
\times\hat c_{r_1,s_1,\uparrow}^\dag(\vk_1) \hat c_{r_2,s_2,\uparrow}\pdag(\vk_2)\hat c_{r_3,s_3,\downarrow}^\dag(\vk_3) \hat c_{r_4,s_4,\downarrow}\pdag(\vk_4)
\end{split}
\end{equation}
the interaction part.

We will assume that there exists some underlying Fermi surface dominating the low-energy physics of the interacting model near half-filling, and that this surface has ``flat parts'' that can be approximated by a straight line segment or {\em Fermi arc} in each nodal region. Furthermore, we assume that the parameter $Q$ is such that each $\vQ_{r,s=\pm}$ lies on this underlying Fermi surface ($Q$ is the analogue of $k_F$ in the corresponding 1D model). We make no assumption on the geometry of the Fermi surface in the antinodal regions. 
 
In the following, we concentrate on that part of \eqref{Kinetic part in eight-flavor representation: main}--\eqref{Interaction part in eight-flavor representation: main} that only involves the nodal fermions ($s=\pm$); the end-result for the effective nodal Hamiltonian is given in the next secion, while the inclusion of antinodal fermions is discussed in Section~\ref{Sect: Integrating out degrees of freedom}. In Appendix~\ref{App:Derivation of the effective model}, the approximations introduced below (except for the continuum limit) are also applied to the antinodal (and in- and out) fermions in order to highlight similarities and differences between the fermions. In the appendices, we also include a nn interaction in the lattice Hamiltonian.

We expand the tight-binding band relations $\epsilon(\vQ_{r,s}+\vk)$ for the nodal fermions as
\begin{equation}
\label{Expansion of nodal tight-binding band relation}
\epsilon(\vQ_{r,s}+\vk)= \epsilon(\vQ_{r,s}) + \varepsilon_{r,s}(\vk) + O(|a\vk|^2), \qquad r,s=\pm
\end{equation}
with
\begin{equation}
\label{Nodal dispersion}
\varepsilon_{r,s}(\vk)= v_F r k_s, \qquad v_F= 2\sqrt{2}\sin(Q)\left[t+2t'\cos(Q)\right]a
\end{equation}
and where we use coordinates $k_\pm = (k_1 \pm k_2)/\sqrt2$. Our first approximation is to only keep terms up to linear order in $|a\vk|$.

The interaction in the Hubbard Hamiltonian consists of those scattering processes $(\vk_2,\vk_4)\to(\vk_1,\vk_3)$ that conserve overall momenta (up to reciprocal lattice vectors). When writing the Hubbard Hamiltonian in terms of the operators $\hat c^{(\dag)}_{r,s,\alpha}(\vk)$, conservation of momenta corresponds to the following requirement
\begin{equation}
\label{Conservation of momenta}
(\vQ_{r_1,s_1} + \vk_1) - (\vQ_{r_2,s_2} + \vk_2) + (\vQ_{r_3,s_3} + \vk_3) - (\vQ_{r_4,s_4} + \vk_4)\in (2\pi/a)\Z^2
\end{equation}
with $\vk_j\in\Lambda_{r_j,s_j}^*$. The next approximation is to reduce the number of (nodal) interaction terms in the Hubbard Hamiltonian by imposing the additional constraint
\begin{equation}
\label{Q-vector constraint}
\vQ_{r_1,s_1} - \vQ_{r_2,s_2} + \vQ_{r_3,s_3} - \vQ_{r_4,s_4}\in (2\pi/a)\Z^2
\end{equation}
for interaction terms that we keep. If all momenta lie strictly on a Fermi arc, the constraint \eqref{Q-vector constraint} follows from momentum conservation. All possible combinations of $(r_j,s_j)$ satisfying this constraint when $Q\neq\pi/2$ are given in Table~\ref{Flavor combinations} in Appendix~\ref{App:Derivation of the effective model}. If $Q=\pi/2$ there are additional (and potentially gap-inducing) umklapp processes; it is tempting to identify this value with the half-filled model. 

Obvious solutions to the constraint in \eqref{Q-vector constraint} is to set either $(r_1,s_1)=(r_2,s_2)$ and $(r_3,s_3)=(r_4,s_4)$, or $(r_1,s_1)=(r_4,s_4)$ and $(r_2,s_2)=(r_3,s_3)$. These combinations naturally lead to the definition of density- and spin operators $\hat \rho_{r,s}\pdag$ and $\hat S_{r,s}^i$, $i=1,2,3$, corresponding to each pair of flavor indices. For example, the nodal density operators are
\begin{equation}
\label{Flavor density operators}
\hat \rho_{r,s}(\vp) =  \sum\limits_{\alpha=\uparrow,\downarrow}\sum\limits_{\vk_1,\vk_2\in\Lambda^*_{r,s}}\hat c^\dag_{r,s,\alpha}(\vk_1)\hat c\pdag_{r,s,\alpha}(\vk_2)\delta_{\vk_1+\vp,\vk_2}.
\end{equation}
The interaction terms in the truncated Hubbard Hamiltonian with the above combinations for $(r_j,s_j)$ are products of these bilinears, i.e.\ $\hat\rho_{r,s}\hat\rho_{r',s'}$ and $\hat{\bf S}_{r,s}\cdot \hat{\bf S}_{r',s'}$. The constraint in \eqref{Q-vector constraint} also allows for interaction terms involving pairing bilinears of the form $\hat\psi^{(\dag)}\hat\psi^{(\dag)}$. We define associated pairing operators denoted by $\hat\Pair_{r,s}^\mu$, $\mu=0,1,2,3$, and write these interaction terms as $\hat \Pair_{r,s}^\dag\cdot\hat \Pair_{r',s'}\pdag$ with $\hat\Pair_{r,s}\pdag = (\hat\Pair^0_{r,s},\hat\Pair^1_{r,s},\hat\Pair^2_{r,s},\hat\Pair^3_{r,s})$. 

The components of the momenta in the nodal sets $\Lambda_{r,s=\pm}^*$ are restricted by cutoffs proportional to the inverse lattice constant. Our partial contnuum limit for the nodal fermions involves removing the cutoff in the directions orthogonal to each Fermi arc. To this end, we normal-order the kinetic part and the bilinears in the truncated interaction with respect to a state $\Omega$ (the Dirac sea) in which all momenta up to the Fermi arcs in the nodal regions are occupied. 

Consider now region $(+,+)$ in Figure~\ref{Partition of BZ}. After removing the cutoff in the $k_+$-direction, it would be possible to bosonize the nodal fermions by treating the $k_+$ as an unbounded 1D chain of momenta and $k_-$ a flavor index labelling each chain. However, as discussed in the introduction, this does not lead to simple bosonic commutation relations for \eqref{Flavor density operators}; only densities with momentum exchange in the $k_+$-direction would behave as bosons and one cannot treat momentum exchange between fermions on different chains. 
Instead, it is more fruitful to first do a Fourier transformation (change of basis) in the $k_-$-direction and then bosonize the fermions using a new flavor index $x_-$ \cite{Luther:1994,Langmann2:2010}. If one also removes the cutoff in the $k_-$-direction, the commutation relations of the (normal-ordered and rescaled) densities in \eqref{Flavor density operators} become that of 2D bosons. However, this limit is delicate as the (normal-ordered) Hamiltonian would then no longer be bounded from below; see the next section. 

A mathematically more sound way to proceed is to keep the cutoff and instead modify the nodal density operators in \eqref{Flavor density operators}; we define the normal-ordered density operators
\begin{equation}
\hat{J}_{r,s=\pm}^0(\vp) = \sum\limits_{\alpha}\sum\limits_{\vk_1,\vk_2\in\Lambda^*_{s}} :\!\hat c^\dag_{r,s,\alpha}(\vk_1)\hat c\pdag_{r,s,\alpha}(\vk_2)\!:\sum\limits_{n\in\Z}\delta_{\vk_1+\vp + 2\pi n\ve_{ - s}/\ta,\vk_2}
\end{equation}
with $\ve_{-s}$ a unit vector in the direction of the Fermi arc. Here $\tilde{a} = \sqrt{2}a/({1-\kappa})$ with the length of each Fermi arc given by $2\pi/\ta$. This operator is obtained from \eqref{Flavor density operators} by adding ``umklapp terms'' corresponding to $n\neq 0$. As shown in \cite{deWoulLangmann2:2010}, it is possible to send $\ta\to 0^+$ on the level of correlation functions. We do a similar regularization for the spin operators. With this, one obtains our effective nodal Hamiltonian; see Equation~\eqref{Regularised nodal Hamiltonian} in the next section.

\section{\label{Sect: Bosonization of nodal fermions}Nodal fermions}
We formulate the nodal part of the effective QFT model obtained from our partial continuum limit of the 2D Hubbard model near half-filling. We also show that the nodal fermions can be bosonized using exact methods. Some of these results are straightforward generalisations of the corresponding ones obtained for the so-called {\em Mattis model} in \cite{deWoulLangmann2:2010}, and in those instances we will be rather brief in the presentation. Further mathematical details are also given in Appendix~\ref{Appendix Bosonization}. In this section, the flavor indices are always $r,s=\pm$.

\subsection{The nodal Hamiltonian}
We rescale the nodal fermion operators by setting $\hat \psi_{r,s,\alpha}(\vk)=L/(2\pi)\hat c_{r,s,\alpha}(\vk)$ such that
\begin{equation}
\label{Fermion flavor anticommutation relations: main}
\{\hat\psi\pdag_{r,s,\alpha}(\vk),\hat\psi^{\dag}_{r',s',\alpha'}(\vk')\} = [L/(2\pi)]^2\delta_{r,r'}\delta_{s,s'}\delta_{\alpha,\alpha'}\delta_{\vk,\vk'}, \quad \{\hat\psi\pdag_{r,s,\alpha}(\vk),\hat\psi\pdag_{r',s',\alpha'}(\vk')\} = 0.
\end{equation}
The momenta $\vk$ are in the (unbounded) sets
\begin{equation}
\label{Continuum nodal momentum region: main}
\Lambda_{s}^* = \left\{\vk\in \frac{2\pi}{L}\Bigl(\mathbb{Z} +\frac{1}{2}\Bigr)^2 \; : \; -\frac{\pi}{\ta} \leq k_{-s}< \frac{\pi}{\ta} \right\}.
\end{equation}
The nodal part of the effective model is obtained from a Dirac vacuum $\Omega$ satisfying
\begin{equation}
\label{Highest Weight Condition: main}
\hat\psi\pdag_{r,s,\alpha}(\vk)\Omega = \hat\psi^\dag_{r,s,\alpha}(-\vk)\Omega= 0, \quad \mbox{for all } \vk\in\Lambda_{s}^*\; \mbox{ such that }\; rk_s>0
\end{equation} 
with $\langle\Omega,\Omega\rangle=1$. The specific choice of filling for the antinodal fermion states in $\Omega$ is unimportant; we assume for simplicity that no state is occupied. We also introduce ordinary fermion normal-ordering with respect to $\Omega$ such that $:\! \cO\!: \, = \cO - \langle \Omega, \cO\Omega\rangle$ for fermion bilinears $\cO=\hat\psi^\dag_{r,s,\alpha}(\vk) \hat\psi\pdag_{r',s',\alpha'}(\vk')$.

We define the following nodal bilinear operators
\begin{align}
\label{Nodal density- and spin operators}
\hat J_{r,s}^\mu(\vp) & = \sum\limits_{\alpha,\beta}\sum\limits_{\vk_1,\vk_2\in\Lambda^*_{s}}\sum\limits_{n\in\Z} \tPiL^2 :\!\hat\psi^\dag_{r,s,\alpha}(\vk_1)\sigma_{\alpha,\beta}^\mu\hat\psi\pdag_{r,s,\beta}(\vk_2)\!:
\delta_{\vk_1+\vp,\vk_2 + 2\pi n\ve_{ - s}/\ta} 
\\
\label{Nodal pairing operators}
\hat \Pair_{r,s}^\mu(\vp) & = \frac12\sum\limits_{\alpha,\beta}\sum\limits_{\vk_1,\vk_2\in\Lambda^*_{r,s}} \tPiL^2 \hat\psi\pdag_{-r,s,\alpha}(\vk_1)\sigma_{\alpha,\beta}^\mu\hat\psi\pdag_{r,s,\beta}(\vk_2)\delta_{\vk_1+\vk_2,\vp}
\end{align}
with $r,s=\pm$ and $\mu=0,1,2,3$; here $\sigma^i$, $i=1,2,3$, are the Pauli matrices, $\sigma_{\alpha,\beta}^0=\delta_{\alpha,\beta}$ and the momenta $\vp$ are in the set
\begin{equation}
\label{Boson momentum set: main}
\tilde\Lambda_{s}^* = \left\{\vp=(p_+,p_-)\in (2\pi/L)\Z² \; : \; -{\pi}/{\ta} \leq p_{-s}< {\pi}/{\ta} \right\}.
\end{equation}
Spin operators are given by the simple rescaling $\hat S_{r,s}^i=\hat J_{r,s}^i/2$. 
We note that removing the cutoff in the summation of momenta $\vk_1,\vk_2$ in \eqref{Nodal pairing operators} would lead to ill-defined operators \cite{CareyRuijsenaars:1987}. For example, acting with such operators on $\Omega$ would result in a state of infinite norm.  

The nodal part of the effective Hamiltonian is now defined as
\begin{equation}
\begin{split}
\label{Regularised nodal Hamiltonian}
&\Hn = \HWZW + U \sum\limits_{r,r',s=\pm}\sum\limits_{\vp}\Bigl(\frac{a}{L}\Bigr)^2\chi(\vp)\hat\Pair_{r,s}^\dag(\vp)\cdot\hat \Pair_{r',-s}\pdag(\vp)\\
&\HWZW = \HWZWKin+\HWZWInt
\end{split}
\end{equation}
with
\begin{equation}
\label{WZW Hamiltonian: Kinetic term}
\HWZWKin = v_F \sum\limits_{\alpha=\pm}\sum\limits_{r,s=\pm} \sum\limits_{\vk\in\Lambda_s^*} \Bigl({\frac{{2\pi}}{L}}\Bigr)^2 rk_s :\hat\psi_{r,s,\alpha }^\dag(\vk)\hat\psi_{r,s,\alpha}\pdag(\vk): 
\end{equation}
the free part, and
\begin{equation}
\begin{split}
\label{WZW Hamiltonian: Interaction term}
\HWZWInt = \frac{U}{2}\sum\limits_{\vp}\Bigl(\frac{a}{L}\Bigr)^2 \chi(\vp)\Bigl( \sum\limits_{s =\pm}\bigl(\sum\limits_{r=\pm}  \hat J_{r,s}^{0\dag} \hat J_{r,s}^0 
+ \hat J_{+,s}^{0\dag}\hat J_{-,s}^0- {\hat {\bf J}}_{+,s}^\dag \cdot {\hat {\bf J}}_{-,s}\pdag\bigr)
\\
+ \sum\limits_{r,r'=\pm} \bigl( \hat J_{r,+}^{0\dag}\hat J_{r',-}^0 - {\hat {\bf J}}_{r,+}^\dag \cdot {\hat {\bf J}}_{r',-}\pdag\bigr)\Bigr)
\end{split} 
\end{equation}
the density- and spin interaction part; here ${\hat {\bf J}}_{r,s}=(\hat J_{r,s}^1,\hat J_{r,s}^2,\hat J_{r,s}^3)$ and we suppress common arguments of $\vp$. Furthermore, we have introduced a cutoff function for possible momentum exchange in the interaction by
\begin{equation}
\label{Simplified cutoff}
\chi(\vp)=\begin{cases}
1& \text{if } - {\pi}/{\ta} \le p_{\pm} < {\pi}/{\ta}\\
0& \text{otherwise}
\end{cases}.
\end{equation}

The nodal Hamiltonian in \eqref{Regularised nodal Hamiltonian} contains different types of scattering processes. Terms involving the bilinears in \eqref{Nodal density- and spin operators} correspond to processes for which both fermions remain near the same Fermi arc, and for which their spin projection may or may not be reversed. In contrast, terms involving \eqref{Nodal pairing operators} are such that both fermions are scattered from one Fermi arc to another. As we will see below, these latter terms cannot be easily analyzed using our methods. 

We also summarize our conventions for Fourier transforms of nodal operators (similar expressions can be found in \cite{deWoulLangmann2:2010}). Define nodal fermion field operators by
\begin{equation}
\label{Position space operators}
\psi_{r,s,\alpha}\pdag(\vx) = \frac1{2\pi} \sum_{\vk\in\Lambda^*_{s}} \tPiL^2  \hat\psi_{r,s,\alpha}(\vk)\ee^{\ii\vk\cdot\vx} \qquad (s=\pm),
\end{equation}
with ``positions'' $\vx$ in 
\begin{equation}
\Lambda_s = \left\{\vx\in\R^2\, : \, x_s\in\mathbb{R},\;  x_{-s}\in \ta\mathbb{Z},\; -{L}/{2}\leq x_\pm < {L}/{2} \right\}
\end{equation}
and which obey the anticommutation relations
\begin{equation}
\label{Nodal fermion anticommutation relations in position space}
\{\psi\pdag_{r,s,\sigma}(\vx), \psi^{\dag}_{r',s',\sigma'}(\vy)\} = \delta_{r,r'}\pdag\delta_{s,s'}\pdag\delta_{\sigma,\sigma'}\pdag\tilde\delta_s\pdag(\vx-\vy), \qquad \tilde\delta_s(\vx) = \delta(x_s)\frac1\ta\delta_{x_{-s},0}. 
\end{equation}
The (regularized) Fourier transforms of the nodal density- and spin operators in \eqref{Nodal density- and spin operators} are defined as
\begin{equation}
\begin{split}
\label{Nodal densities in position space}
J_{r,s}^\mu(\vx;\epsilon) = \sum_{\vp\in \tilde\Lambda^*_{s}}\frac1{L^2}\hat{J}_{r,s}^\mu(\vp)\ee^{\ii\vp\cdot\vx-\epsilon |p_s|/2}, \qquad J_{r,s}^\mu(\vx)=\lim_{\epsilon\to 0^+} J_{r,s}^\mu(\vx;\epsilon)
\end{split}
\end{equation}
with $\epsilon>0$ infinitesimal. Using these operators, it is for example possible to rewrite $H_0$ in \eqref{WZW Hamiltonian: Kinetic term} in ``position'' space and thus obtain a well-defined regularised expression replacing the free part of \eqref{WZW Hamiltonian in position space}
\begin{equation}
\label{Regularized free nodal Hamiltonian in position space}
\HWZWKin = v_F\sum\limits_{\alpha=\pm}\sum\limits_{r,s=\pm}\tintUnder{s}\ud^2 x\, :\! \psi^\dag_{r,s,\alpha}(\vx)(-\ii r\partial_s)\psi_{r,s,\alpha}\pdag(\vx)\!:
\end{equation}
where we have defined
\begin{equation}
\label{tint}
\tintUnder{s}\ud^2x\, = \int\limits_{-L/2}^{L/2}\ud x_s \sum_{x_{-s}\in\Lambda_{1\rm D}}\ta
\end{equation}
and
\begin{equation}
\label{1D lattice: main}
\Lambda_{1\rm D} = \left\{ x \in \ta\mathbb{Z}\; : \;  -{L}/{2}\leq x <{L}/{2} \right\}.
\end{equation}
The free part of the nodal Hamiltonian \eqref{Regularized free nodal Hamiltonian in position space} can thus be interpreted as a system of 1D (massless) Dirac fermions where, for each $s=\pm$, $x_s\in\bigl[-L/2,L/2\bigr)$ is a (continuous) spatial variable and $x_{-s}\in\Lambda_{1D}$ a (discrete) flavor index. Analogous regularized expressions may be written down for the interaction part of \eqref{WZW Hamiltonian in position space}. Likewise, insertion of the inverse relation to \eqref{Position space operators} into \eqref{Nodal densities in position space}, using \eqref{Nodal density- and spin operators}, yields \eqref{Current operators in position space}.

\subsection{\label{Sec: Bosonization}Bosonization}
The presence of the Dirac vacuum satisfying \eqref{Highest Weight Condition: main} leads to anomalous commutator relations \cite{MattisLieb:1965} for the fermion bilinears in \eqref{Nodal density- and spin operators} (see Appendix~\ref{Appendix Bosonization} for proof)
\begin{equation}
\label{Current algebra: main}
\begin{split}
\left[\hat J_{r,s}^0(\vp),\hat J_{r,s}^0(\vp')\right] & =  r\frac{4\pi p_s}{\ta}\Bigl(\frac{L}{2\pi}\Bigr)^2 \delta_{\vp + \vp',\vzero}
\\
\left[\hat J_{r,s}^i(\vp),\hat J_{r,s}^j(\vp')\right] & = 2\ii\sum\limits_{k=1}^3 \epsilon_{ijk}\hat J_{r,s}^k(\vp + \vp') 
+  r\frac{4\pi p_s}{\ta}\Bigl(\frac{L}{2\pi}\Bigr)^2 \delta_{\vp + \vp',\vzero}\delta_{i,j}
\end{split}
\end{equation}
with all other commutators vanishing; $\epsilon_{ijk}$ is the totally antisymmetric tensor and $\epsilon_{123}=1$. Furthermore, $\hat J_{r,s}^\mu(\vp)\Omega = 0$ for all $\vp$ such that $rp_s\geq 0$. Using \eqref{Current algebra: main} together with \eqref{Nodal densities in position space}, one obtains the commutation relations in \eqref{WZW commutation relations in position space} (everywhere replacing $\delta(\vx)$ with $\tilde\delta_s(\vx)$ defined in \eqref{Nodal fermion anticommutation relations in position space}). 

We introduce spin-dependent densities,
\begin{equation}
\label{Nodal spin-density operators}
\hat J_{r,s,\uparrow}(\vp) = \bigl(\hat J_{r,s}^0(\vp) + \hat J_{r,s}^3(\vp)\bigr)/2, \qquad \hat J_{r,s,\downarrow}(\vp) = \bigl(\hat J_{r,s}^0(\vp) - \hat J_{r,s}^3(\vp)\bigr)/2,
\end{equation}
which by \eqref{Current algebra: main} satisfy the commutation relations
\begin{equation} 
\label{Commutation relations of densities}
\Bigl[\hat{J}_{r,s,\alpha}(\vp), \hat{J}_{r,s,\alpha'}(\vp')\Bigr]= r \delta_{\alpha,\alpha'} \frac{2\pi p_s }{\ta}\Bigl(\frac{L}{2\pi}\Bigr)^2\delta_{\vp+\vp',\vzero}.
\end{equation}
It follows that the rescaled densities
\begin{equation}
\label{Boson operators from nodal density operators}
b_{s,\alpha}(\vp) =
\begin{cases}
-\frac{\ii}L\sqrt{\frac{2\pi\ta}{|p_s|}} \hat{J}_{+,s,\alpha}(\vp) & \text{if $p_s>0$}\\
\phantom{-}\frac{\ii}L\sqrt{\frac{2\pi\ta}{|p_s|}}\hat{J}_{-,s,\alpha}(\vp) & \text{if $p_s<0$}
\end{cases}
\end{equation} 
obey the defining relations of 2D boson creation- and annihilation operators,
\begin{equation}
[b\pdag_{s,\alpha}(\vp),b^\dag_{s',\alpha'}(\vp')]=\delta_{s,s'}\delta_{\alpha,\alpha'}\delta_{\vp,\vp'}, \qquad [b\pdag_{s,\alpha}(\vp),b\pdag_{s',\alpha'}(\vp')]=0, \qquad b\pdag_{s,\alpha}(\vp)\Omega=0.
\end{equation}
The boson operators in \eqref{Boson operators from nodal density operators} are defined for momenta $\vp\in\tilde\Lambda_{s}^*$ such that $p_s\neq 0$; we denote this set as $\hat\Lambda_{s}^*$ (see \eqref{Momentum set with zero-component}). Corresponding to momenta with $p_s=0$, we also introduce so-called zero mode operators, or simply {\em zero modes}, $N_{r,s,\alpha}(x)$ with $x\in\Lambda_{1\rm D}$ (see \eqref{1D lattice: main}); their definition is given in Appendix~\ref{Appendix Bosonization}. To complete the bosonization of the nodal fermions, we also need the so-called {\em Klein factors} $R_{r,s,\alpha}(x)$ conjugate to the zero modes. These are sometimes called {\em charge shift-} or {\em ladder operators} \cite{Haldane:1979} as they raise or lower the number of fermions (with flavor indices $(r,s,\alpha,x)$) by one when acting on the Dirac vacuum. 
The Klein factors, together with the boson operators introduced above, span the nodal part of the fermion Fock space when acting on the Dirac vacuum; see Appendix~\ref{Appendix Bosonization} for details. This enables us to express nodal operators in terms of Klein factors and density operators; in particular, the fermion field operator in \eqref{Position space operators} has the form
\begin{equation}
\begin{split} 
\psi_{r,s,\alpha}\pdag(\vx) \sim \frac1{\sqrt{2\pi \ta\epsilon}} R_{r,s,\alpha}(x_{-s})^{-r} \exp\Bigl(r \frac{\ta}{2\pi} \sum_{\vp\in\hat\Lambda_{s}}\tPiL^2 \frac1{p_s} \hat{J}_{r,s,\alpha}(\vp)\ee^{\ii \vp\cdot\vx}\ee^{-\epsilon |p_s|/2} \Bigr) 
\end{split}
\end{equation}
with $\epsilon\to 0^+$ implicit; precise statements are given in Appendix~\ref{Appendix Bosonization}. 

We define {\em boson normal-ordering} with respect to the Dirac vacuum $\Omega$ such that 
\begin{equation}
\label{Boson normal-ordering}
\xxa \hat{J}_{r,s}^\mu(\vp)\hat{J}_{r',s'}^\mu(\vp')\xxe \,\define 
\begin{cases}
\hat{J}_{r,s}^\mu(\vp)\hat{J}_{r',s'}^\mu(\vp') & \text{if $rp_s<0$}\\
\hat{J}_{r',s'}^\mu(\vp')\hat{J}_{r,s}^\mu(\vp) & \text{if $rp_s\geq0$}
\end{cases} \qquad (\mu=0,1,2,3)
\end{equation}
(analogous expressions hold for $\hat{J}_{r,s,\alpha}$). Then the following operator identities hold true
\begin{equation}
\label{Sugawara construction: main}
\begin{split}
\sum\limits_{\alpha=\uparrow,\downarrow}\sum\limits_{\vk\in\Lambda_s^*} \Bigl({\frac{{2\pi }}{L}}\Bigr)^2 rk_s :\!\hat\psi_{r,s,\alpha }^\dag\hat\psi_{r,s,\alpha}\pdag\!: & = \ta\pi \sum\limits_{\alpha=\uparrow,\downarrow}\sum_{\vp\in\tilde\Lambda_s^*} \frac{1}{L^2} \xxa \hat{J}_{r,s,\alpha}^\dag \hat{J}_{r,s,\alpha}\pdag\xxe
\\
& = \frac{\ta\pi}{2}\sum\limits_{\vp\in \tilde \Lambda_s^*} \frac{1}{L^2} \xxa\Bigl(\hat J_{r,s}^{0\dag}\hat J_{r,s}^0 + \frac{1}{3}{\hat{\bf J}}_{r,s}^\dag \cdot {\hat{\bf J}}_{r,s}\pdag \Bigr)\xxe 
\end{split}
\end{equation}
with the momentum sets defined in \eqref{Continuum nodal momentum region: main} and \eqref{Boson momentum set: main}. The first identity is an application of the Kronig identity, while the second is a Sugawara construction; see Appendix~\ref{Appendix Bosonization}.

\subsection{\label{Sec: An exactly solvable model of 2D electrons}An exactly solvable model of 2D electrons}
We discuss the bosonization of the nodal Hamiltonian in \eqref{Regularised nodal Hamiltonian} using the results obtained above. Inserting the last expression of \eqref{Sugawara construction: main} into \eqref{WZW Hamiltonian: Kinetic term} gives \eqref{Spin-charge separation in position space} with (cf. \eqref{Spin and charge parts in position space})
\begin{align}
\label{Charge part of nodal Hamiltonian: main}
\begin{split}
\HWZW_\Charge\pdag = \frac{v_F\pi\ta }{2}\sum\limits_{\vp} \frac{1}{L^2}\xxa\Big(\sum\limits_{r,s=\pm}\left(\left(1+ 2\gamma\chi(\vp) \right)\hat J_{r,s}^{0\dag} {\hat J}_{r,s}^0 + \gamma\chi(\vp)\hat J_{r,s}^{0\dag} {\hat J}_{-r,s}^0 \right)  
\\ 
+ 2\gamma\chi(\vp)\sum\limits_{r,r'=\pm}\hat J_{r,+}^{0\dag}{\hat J}_{r',-}^0 \Bigr) \xxe
\end{split}
\\
\label{Spin part of nodal Hamiltonian: main}
\begin{split}
\HWZW_\Spin\pdag =\frac{v_F\pi\ta }{2} \sum\limits_{\vp} \frac{1}{L^2}\xxa\Big( \sum\limits_{r,s=\pm} \left(\hat {\bf J}_{r,s}^\dag  \cdot {\hat {\bf J}}_{r,s}\pdag/3 - \gamma{\chi}(\vp)\hat {\bf J}_{r,s}^\dag\cdot {\hat {\bf J}}_{-r,s}\pdag \right)   
\\ 
- 2\gamma{\chi}(\vp) \sum\limits_{r,r'=\pm}\hat {\bf J}_{r,+}^\dag\cdot{\hat {\bf J}}_{r',-}\pdag\Bigr) \xxe
\end{split}
\end{align}
and with the dimensionsless coupling constant
\begin{equation}
\gamma = \frac{a^2 U}{2\pi\ta v_F }.
\end{equation}
We emphasize that this does not imply (exact) spin-charge separation of the nodal Hamiltonian; there is also the second term on the right hand side of \eqref{Regularised nodal Hamiltonian} that does not have a simple bosonized form (although it can indeed be expressed in terms of Klein factors, density- and spin operators using Proposition~\ref{Proposition Fermions from Bosons} in Appendix~\ref{Appendix Bosonization}). 

A complete analysis of \eqref{Charge part of nodal Hamiltonian: main}--\eqref{Spin part of nodal Hamiltonian: main} will not be attempted in the present paper. Instead, we will focus in the remainder of this section on the ``abelian'' part of $H$ obtained by breaking manifest spin rotation invariance, and which we denote by $\HMattis$ due to its similarity to the so-called Mattis Hamiltonian in \cite{deWoulLangmann2:2010}. More specifically, we write
\begin{equation}
\label{Break manifest SU(2): main}
\begin{split}
\HWZW = \HMattis -\frac{U}{4}\sum\limits_{\vp}\Bigl(\frac{a}{L}\Bigr)^2 \chi(\vp)\Bigl( &\sum\limits_{s =\pm}\bigl(\hat J_{+,s}^+(-\vp) \hat J_{-,s}^-(\vp)+h.c.\bigr)
\\
&+ \sum\limits_{r,r'=\pm} \bigl(  \hat J_{r,+}^+(-\vp)\hat J_{r',-}^-(\vp)+h.c.\bigr)\Bigr) 
\end{split}
\end{equation}
with the raising- and lowering operators defined as usual, $\hat J_{r,s}^\pm = \bigl(\hat J_{r,s}^1 \pm \ii \hat J_{r,s}^2\bigr)/2$, and where $\HMattis$ only depends on $\hat J_{r,s}^0$ and $\hat J_{r,s}^3$. Using results from Section~\ref{Sec: Bosonization}, it is possible to write the Hamiltonian $\HMattis$ in terms of free bosons. Define
\begin{gather}
\begin{aligned}
\label{hatPihatPhi}
\hat\Phi_{\Charge;s}(\vp)\define&\sqrt{\frac{\ta}{8\pi}}\frac1{\ii p_s}\Bigl(\hat{J}_{+,s}^0(\vp) + \hat{J}_{-,s}^0(\vp) \Bigr), & \quad  \hat\Pi_{\Charge;s}(\vp)\define&\sqrt{\frac{\ta}{8\pi}}\Bigl(-\hat{J}_{+,s}^0(\vp) + \hat{J}_{-,s}^0(\vp) \Bigr) \\ 
\hat\Phi_{\SpinZ;s}(\vp)\define&\sqrt{\frac{\ta}{8\pi}}\frac1{\ii p_s}\Bigl(\hat{J}_{+,s}^3(\vp) + \hat{J}_{-,s}^3(\vp) \Bigr), & \quad  \hat\Pi_{\SpinZ;s}(\vp)\define&\sqrt{\frac{\ta}{8\pi}}\Bigl(-\hat{J}_{+,s}^3(\vp) + \hat{J}_{-,s}^3(\vp) \Bigr)
\end{aligned} 
\end{gather} 
for $s=\pm$ and 2D momenta $\vp\in\hat\Lambda_s^*$. It follows that these obey the defining relations of 2D neutral bosons, i.e.\ 
\begin{equation}
{[}\hat\Phi\pdag_{X;s}(\vp),\hat\Pi^\dag_{X';s'}(\vp'){]}=\ii\delta_{X,X'}\delta_{s,s'}\Bigl(\frac{L}{2\pi}\Bigr)^2\delta_{\vp,\vp'}
\end{equation} 
(all other commutators vanishing) and
\begin{equation}
\hat\Pi^\dag_{X;s}(\vp)=\hat\Pi\pdag_{X;s}(-\vp),\qquad \hat\Phi^\dag_{X;s}(\vp)=\hat\Phi\pdag_{X;s}(-\vp),
\end{equation}    
where we have used the symbolic notation $X,X'=\Charge,\SpinZ$. Furthermore, applying the first equality in \eqref{Sugawara construction: main} to \eqref{WZW Hamiltonian: Kinetic term}, together with \eqref{hatPihatPhi}, allows us to write
\begin{equation}
\label{Spinfull Mattis Hamiltonian: main}
\HMattis = \HWZW_\Charge\pdag + \HWZW_\SpinZ\pdag
\end{equation}
with
\begin{align}
\label{Charge part of spinfull Mattis Hamiltonian: main}
\begin{split}
\HWZW_\Charge\pdag = &\frac{v_F}{2}\sum_{s=\pm}\sum_{\vp\in\hat\Lambda^*_s}\tPiL^2 \xxa\Bigl(\bigl(1+\gamma\chi(\vp)\bigr)\hat\Pi^\dag_{\Charge;s}\hat\Pi\pdag_{\Charge;s}
\\
&+ \bigl(1+3\gamma\chi(\vp)\bigr)p_s^2\hat\Phi^\dag_{\Charge;s}\hat\Phi\pdag_{\Charge;s}+2\gamma p_+p_-\chi(\vp) \hat\Phi^\dag_{\Charge;s}\hat\Phi\pdag_{\Charge;-s}\Bigr)\xxe + z.m.
\end{split}
\\
\label{Spin part of spinfull Mattis Hamiltonian: main}
\begin{split}
\HWZW_\SpinZ = &\frac{v_F}{2}\sum_{s=\pm}\sum_{\vp\in\hat\Lambda^*_s}\tPiL^2 \xxa\Bigl(\bigl(1+\gamma\chi(\vp)\bigr)\hat\Pi^\dag_{\SpinZ;s}\hat\Pi\pdag_{\SpinZ;s} 
\\
&+ \bigl(1-\gamma\chi(\vp)\bigr)p_s^2\hat\Phi^\dag_{\SpinZ;s}\hat\Phi\pdag_{\SpinZ;s}-2\gamma p_+p_-\chi(\vp) \hat\Phi^\dag_{\SpinZ;s}\hat\Phi\pdag_{\SpinZ;-s}\Bigr)\xxe + z.m.
\end{split}
\end{align}
with $z.m.$ denoting terms involving zero mode operators; a complete solution including the zero modes is given in Appendix~\ref{Appendix: Bosonization of the nodal Hamiltonian}. 

The charge- and spin parts of $\HMattis$ in \eqref{Spinfull Mattis Hamiltonian: main} each have the same structure as the bosonized Hamiltonian of the so-called {\em Mattis model} of spinless fermions studied in \cite{deWoulLangmann2:2010} (compare Equation~(3.3) in \cite{deWoulLangmann2:2010} with \eqref{Charge part of spinfull Mattis Hamiltonian: main} and \eqref{Spin part of spinfull Mattis Hamiltonian: main}). As for the Mattis Hamiltonian, the right hand side of \eqref{Spinfull Mattis Hamiltonian: main} can be diagonalised by a Bogoliubov transformation into a sum of decoupled harmonic oscillators and zero mode terms. To this end, define
\begin{equation}
b_{\Charge;s}(\vp) = \bigl(b_{s,\uparrow}(\vp)+b_{s,\downarrow}(\vp)\bigr)/\sqrt{2}, \qquad b_{\SpinZ;s}(\vp) = \bigl(b_{s,\uparrow}(\vp)-b_{s,\downarrow}(\vp)\bigr)/\sqrt{2}.
\end{equation}
The Hamiltonian in \eqref{Spinfull Mattis Hamiltonian: main} can then be diagonalized by a unitary operator $\cU$ as follows (see Theorem~\ref{Theorem: Diagonalisation of Spinfull Mattis Hamiltonian} in Appendix~\ref{Appendix: Bosonization of the nodal Hamiltonian})
\begin{equation}
\cU^\dag \HMattis\cU = \sum_{s=\pm}\sum_{\vp\in\hat\Lambda_s^*}\left(\omega_{\Charge;s}\pdag(\vp) b_{\Charge;s}^\dag(\vp)b\pdag_{\Charge;s}(\vp) + \omega_{\SpinZ;s}\pdag(\vp) b_{\SpinZ;s}^\dag(\vp)b\pdag_{\SpinZ;s}(\vp)\right) +\cE^{(0)} +z.m.
\end{equation}
with 
\begin{equation}
\label{Charge dispersion relation: main}
\omega_{\Charge;\pm}\pdag(\vp)= \begin{cases} \tilde{v}_{F}^\Charge\sqrt{\frac12\Bigl( |\vp|^2 \pm \sqrt{|\vp|^4-A_\Charge\bigl(2p_+p_-\bigr)^2 }\; \Bigr)}& \mbox{ if }\; \gamma\chi(\vp)p_+p_-\neq 0
 \\v_F\sqrt{\bigl(1+2\gamma\chi(\vp)\bigr)^2-\bigl(\gamma\chi(\vp)\bigr)^2}|p_\pm|& \mbox{ if }\;\gamma\chi(\vp)p_+p_-= 0  \end{cases}
\end{equation} 
\begin{equation}
\label{Charge AvF: main} 
A_\Charge= 1-\bigl[{2\gamma}/({1+3\gamma})\bigr]^2  ,\qquad \tilde{v}_F^\Charge= v_F \sqrt{\bigl(1+2\gamma\bigr)^2-\gamma^2}
\end{equation} 
and
\begin{equation}
\label{Spin dispersion relation: main}
\omega_{\SpinZ;\pm}\pdag(\vp)= \begin{cases} \tilde{v}_{F}^\SpinZ\sqrt{\frac12\Bigl( |\vp|^2 \pm \sqrt{|\vp|^4-A_\SpinZ\bigl(2p_+p_-\bigr)^2 }\; \Bigr)}& \mbox{ if }\; \gamma\chi(\vp)p_+p_-\neq 0
 \\v_F\sqrt{1-\bigl(\gamma\chi(\vp)\bigr)^2}|p_\pm|& \mbox{ if }\;\gamma\chi(\vp)p_+p_-= 0  \end{cases}
\end{equation} 
\begin{equation}
\label{Spin AvF: main} 
A_\SpinZ= 1-\bigl[2{\gamma}/({1 -\gamma})\bigr]^2  ,\qquad \tilde{v}_F^\SpinZ= v_F \sqrt{1-\gamma^2}
\end{equation} 
the boson dispersion relations, and
\begin{equation}
\cE^{(0)} = \frac12\sum_{s=\pm}\sum_{\vp\in\hat\Lambda_s^*} \bigl(\omega_{\Charge;s}\pdag(\vp)+\omega_{\SpinZ;s}\pdag(\vp)-2v_F |p_s|\bigr)
\end{equation}
the groundstate energy of $\HMattis$. This is well-defined if \eqref{Constraint on gamma} is fulfilled. For the special case $t'=0$, $\kappa=1/2$, and $Q\to\pi/2$, one obtains the upper bound $U/t<16\pi/3$.

In principle, one can now obtain the complete solution for the model defined by $\HMattis$ by stepwise generalizing the results given in \cite{deWoulLangmann2:2010} to the present case. For example, all correlation functions of nodal fermion operators in the thermal equilibrium state obtained from $\HMattis$ can be computed exactly by analytical methods. 
Furthermore, as shown in \cite{deWoulLangmann2:2010}, zero modes do not contribute to correlation functions in the thermodynamic limit $L\to\infty$ (much like in 1D). The only exception are the Klein factors that need to be handled with some care; see Section~3.3 in \cite{deWoulLangmann2:2010}.

\section{\label{Sect: Integrating out degrees of freedom}Integrating out degrees of freedom}

Up to now, we have studied the part that involves only nodal fermions in the effective Hamiltonian \eqref{Introduction: Effective Hamiltonian near half-filling}. Below, we will propose different ways of also including the antinodal fermions in the analysis. 

\subsection{Integrating out nodal fermions}

The nodal- and antinodal fermions couple through various types of scattering processes in the Hubbard interaction \eqref{Interaction part in eight-flavor representation: main} that we cannot treat in full generality. A simple approximation is to also introduce the constraint in \eqref{Q-vector constraint} for nodal-antinodal processes. This leads to an effective interaction involving nodal- and antinodal bilinears of the same form as in \eqref{Regularised nodal Hamiltonian}; we refer to Appendix~\ref{App:Derivation of the effective model} for details. If we truncate this interaction further by only keeping terms involving the nodal bilinears $J^0_{r,s}$ and $J^3_{r,s}$, it is possible to integrate out the bosonized nodal fermions using a functional integral representation of the partition function. We set (cf.\ \eqref{Regularised nodal-antinodal interaction}; note the abuse of notation for the left hand side)
\begin{equation}
\label{Density- and spin part of nodal-antinodal interaction: main}
H_{na} = \frac{U}{2}\sum\limits_{r,r',s=\pm}\sum\limits_{\vp}\Bigl(\frac{a}{L}\Bigr)^2\chi(\vp) \Bigl(\hat J_{r,s}^{0}(-\vp) \hat J_{r',0}^0(\vp) - {\hat J}_{r,s}^3(-\vp) {\hat J}_{r',0}^3(\vp)\Bigr)
\end{equation}
with the antinodal bilinears ($\mu=0,1,2,3$)
\begin{equation}
\hat J_{r,0}^{\mu}(\vp) =  \sum\limits_{\alpha,\beta=\uparrow,\downarrow}\sum\limits_{\vk_1,\vk_2\in\Lambda^*_{0}} \hat c^\dag_{r,0,\alpha}(\vk_1)\sigma_{\alpha,\beta}^\mu\hat c\pdag_{r,0,\beta}(\vk_2)\delta_{\vk_1+\vp,\vk_2} 
\end{equation} 
and we can write $\Lambda^*_{0}=\Lambda^*_{r,0}$ for the antinodal momenta. Using \eqref{hatPihatPhi},  
\begin{equation}
\label{Nodal-antinodal interaction: main}
\begin{split}
H_{na} = \frac{U}{2}\sqrt{\frac{2}{\pi\ta}}\sum\limits_{r,s=\pm} \sum\limits_{\vp\in\hat\Lambda_s^*} \Bigl(\frac{a}{L}\Bigr)^2 2\pi \ii p_s \chi(\vp)\Bigl( \hat J_{r,0}^{0}(-\vp)\hat\Phi_{\Charge;s}\pdag(\vp) -
\hat J_{r,0}^{3}(-\vp)\hat \Phi_{\SpinZ;s}\pdag(\vp) \Bigr) + z.m. 
\end{split}
\end{equation} 
where $z.m.$ denote terms involving zero mode operators; we will assume throughout this section that their contribution to the functional integral becomes irrelevant in the thermodynamic limit $L\to\infty$. 

The functional integration of the nodal bosons is done exactly as in \cite{Langmann2:2010} (see Section~6.3 and Appendix~C) with the only difference that we now have two independent boson fields instead of one. Performing the (Gaussian) integrals for the fields $\hat\Pi\pdag_{\Charge;s}(\tau,\vp)$ and $\hat\Pi\pdag_{\SpinZ;s}(\tau,\vp)$ yields an action that is at most quadratic in $\hat\Phi\pdag_{\Charge;s}(\tau,\vp)$ and $\hat\Phi\pdag_{\SpinZ;s}(\tau,\vp)$. The interaction between the nodal boson fields and the antinodal fermions, which is linear in the former, can then be removed by completing a square. 
This leads to the induced action
\begin{equation}
\label{Induced action from abelian treatment: main}
S_{a}' = \sum\limits_{n\in\Z}\sum\limits_{r,r'=\pm}\sum\limits_{\vp}\frac{1}{L^2}\left(\hat v_\Charge\pdag(\omega_n,\vp)\hat J_{r,0}^{0\dag}\hat J_{r',0}^0 + \hat v_\SpinZ\pdag(\omega_n,\vp) \hat J_{r,0}^{3\dag} \hat J_{r',0}^3\right) 
\end{equation}
contributing to the full antinodal action; we write $\hat J_{r,0}^{\mu\dag}= \hat J_{r,0}^\mu (-\omega_n,-\vp)$ with boson Matsubara frequencies $\omega_n= 2\pi n/\beta$. The induced density-density interaction potential is found to be 
\begin{equation}
\label{Induced charge interaction potential: main}
\hat v_\Charge\pdag(\omega_n,\vp) = -\frac{a^4U^2}{8\pi\ta v_F}\sum\limits_{s=\pm}\frac{W_{\Charge;s}\pdag(\vp)}{\omega_n^2 + \omega_{\Charge;s}\pdag(\vp)^2} \chi(\vp)
\end{equation}
with
\begin{equation}
\label{W charge: main}
W_{\Charge;\pm}\pdag(\vp) = v_F^2\left(1 + \gamma\right)\Biggl(\left|\vp\right|^2 \pm\frac{\left(p_+^2 - p_-^2\right)^2 + \sqrt{1-A_\Charge}\left(2p_+p_-\right)^2}{\sqrt{\left|\vp\right|^4 - A_\Charge \left(2p_+p_-\right)^2}}\Biggr)
\end{equation}
(see also definitions \eqref{Charge dispersion relation: main}--\eqref{Charge AvF: main}). Likewise, the induced spin-spin interaction potential is
\begin{equation}
\label{Induced spinZ interaction potential: main}
\hat v_\SpinZ\pdag(\omega_n,\vp) = -\frac{a^4U^2}{8\pi\ta v_F}\sum\limits_{s=\pm}\frac{W_{\SpinZ;s}\pdag(\vp)}{\omega_n^2 + \omega_{\SpinZ;s}\pdag(\vp)^2}\chi(\vp)
\end{equation}
with (see \eqref{Spin dispersion relation: main}--\eqref{Spin AvF: main})
\begin{equation}
\label{W spinZ: main}
W_{\SpinZ;\pm}\pdag(\vp) = v_F^2\left(1 - \gamma\right)\Biggl(\left|\vp\right|^2 \pm\frac{\left(p_+^2 - p_-^2\right)^2 - \sqrt{1-A_\SpinZ}\left(2p_+p_-\right)^2}{\sqrt{\left|\vp\right|^4 - A_\SpinZ \left(2p_+p_-\right)^2}}\Biggr).
\end{equation}
We note that the functional form of the induced potentials \eqref{Induced charge interaction potential: main} and \eqref{Induced spinZ interaction potential: main} are both identical to the induced potential found for the spinless model; cf. Equation~(86) in \cite{Langmann2:2010}. 

Furthermore, in the derivation above, we have been rather nonchalant in treating different momentum domains. In particular, in \eqref{Induced action from abelian treatment: main}{\em ff} one should in principle be more careful to distinguish between different cases when components of $\vp$ are zero or not.  We assume this becomes irrelevant in the thermodynamic limit.

The analysis above leads to an effective antinodal action that breaks spin rotation invariance. It is possible to go beyond this abelian treatment by recalling the commutation relations in \eqref{WZW commutation relations in position space}. Rescaling the nodal operators $\tilde J_{r,s}^i(\vx) \define \sqrt{\ta} J_{r,s}^i(\vx)$, one sees that the first term on the right hand side of the commutator
\begin{equation}
\label{Rescaled spin commutation relations}
\left[\tilde J_{r,s}^i(\vx),\tilde  J_{r,s}^j(\vy)\right]  = 2\ii\sqrt{\ta}\sum\limits_{k} {\epsilon_{ijk}\tilde  J_{r,s}^k (\vx)}\delta_s\left(\vx - \vy\right)
+ r\frac{1}{\pi\ii}\delta_{i,j}\partial_s\delta_s\left(\vx - \vy\right)
\end{equation} 
is of lower order in $\ta$ as compared to the second term. This suggest, at least within the functional integral framework, to treat the three components approximately as mutually commuting (bosonic) fields; thus being able to integrate out the nodal fermions while still preserving spin rotation invariance. Results are given in Appendix~\ref{App: Functional integration of nodal bosons}.

\subsection{Integrating out antinodal fermions}

Another interesting possibility is to integrate out the antinodal fermions and obtain an effective action involving only nodal fermions. To do this in a systematic manner, it is useful to return to the representation \eqref{Kinetic part in eight-flavor representation: main}--\eqref{Interaction part in eight-flavor representation: main} of the Hubbard Hamiltonian. The corresponding action for the pure antinodal part can then be written ($\vp\in(2\pi/L)\Z^2$, $-\pi/a\leq p_{1,2}<\pi/a$)
\begin{equation}
\begin{split}
S_a = &\sum_{\alpha=\uparrow,\downarrow}\sum_{r=\pm}\sum_{\vk\in\Lambda^*_0}\int_{0}^\beta \ud \tau\,\bar \psi_{r,0,\alpha}(\tau,\vk)\left(\partial_\tau+\epsilon(\vQ_{r,s}+\vk)-\mu\right) \psi_{r,0,\alpha}\pdag(\tau,\vk)
\\
&+\frac{U}{4}\sum_{r_j=\pm}\sum_{\vp}\Bigl(\frac{a}{L}\Bigr)^2\int_0^\beta \ud\tau \left( \rho_{r_1r_2}^a( \tau ,\vp)\rho_{r_3r_4}^a( \tau ,-\vp) - J_{r_1r_2}^{3,a}( \tau ,\vp)J_{r_3r_4}^{3,a}( \tau ,-\vp) \right)
\end{split}
\end{equation}
with Grassmann fields $\psi_{r,0,\alpha}(\tau,\vk)$, Matsubara time $\tau\in[0,\beta)$ and bilinears
\begin{equation}
\begin{split}
&\rho_{r_1r_2}^a(\tau ,\vp) = \sum\limits_\alpha  {\sum\limits_{\vn \in {\Z^2}} {\sum\limits_{{\vk_1}{\vk_2} \in \Lambda _0^ * } {{\delta _{{\vQ_{{r_1},0}} - {\vQ_{{r_2},0}} + {\vk_1} - {\vk_2},\vp + 2\pi \vn/a}}\bar \psi _{{r_1},0,\alpha }^\dag (\tau ,{\vk_1}){\psi _{{r_2},0,\alpha }}(\tau ,{\vk_2})} } }
\\
&J_{{r_1}{r_2}}^{3,a}(\tau ,\vp) = \sum\limits_{\alpha \alpha '} {\sum\limits_{\vn \in {^2}} {\sum\limits_{{\vk_1}{\vk_2} \in \Lambda _0^ * } {{\delta _{{\vQ_{{r_1},0}} - {\vQ_{{r_2},0}} + {\vk_1} - {\vk_2},\vp + 2\pi \vn/a}}\bar \psi _{{r_1},0,\alpha }^\dag (\tau ,{\vk_1})\sigma _{\alpha \alpha '}^3{\psi _{{r_2},0,\alpha '}}(\tau ,{\vk_2})} } }
\end{split}. 
\end{equation}
The nodal action $S_n$ has the same form with corresponding bilinears $\rho_{r_1r_2}^n(\tau ,\vp)$ and $J_{{r_1}{r_2}}^{3,n}(\tau ,\vp)$. The action for the nodal-antinodal interaction in \eqref{Interaction part in eight-flavor representation: main} has a more complicated form. A simple approximation is to truncate it by only keeping the terms
\begin{equation}
S_{na} = \frac{U}{2}\sum_{r_j=\pm}\sum_{\vp}\Bigl(\frac{a}{L}\Bigr)^2\int_0^\beta \ud\tau\left( \rho_{r_1r_2}^a( \tau ,\vp)\rho_{r_3r_4}^n( \tau ,-\vp) - J_{r_1r_2}^{3,a}( \tau ,\vp)J_{r_3r_4}^{3,n}( \tau ,-\vp) \right).
\end{equation} 
We define the full action as $S=S_n+S_a+S_{na}$. The interaction terms can then be decoupled by introducing two Hubbard-Stratonovich (HS) fields $\phi_0(\tau,\vp)$ and $\phi_S(\tau,\vp)$ such that
\begin{equation}
\label{Action with HS fields}
\begin{split}
S = \sum\limits_{\alpha} {\sum\limits_{r =\pm } {\sum\limits_{s = 0, \pm } {\sum\limits_{\vk \in \Lambda _{r,s}^ *  } {\int\limits_0^\beta  {\ud\tau \,\bar \psi _{r,s,\alpha } \left( {\partial _\tau   + \varepsilon (\vQ_{r,s}  + \vk) - \mu } \right)\psi _{r,s,\alpha } } } } } }  + \frac{U}{4}\sum\limits_{\vp} \Bigl( {\frac{a}{L}} \Bigr)^2\int\limits_0^\beta  \ud\tau
\\ 
 \times\Bigl( {\hat \phi _0^\dag  \hat \phi_0\pdag  + \hat \phi_S^\dag  \hat \phi_S\pdag  - 2\sum\limits_{r_1 r_2} {\bigl( {\ii\hat \phi_0^\dag  ( {\rho _{r_1 r_2 }^{a}  + \rho _{r_1 r_2 }^n } ) + \hat \phi _S^\dag  ( {J_{r_1 r_2 }^{3,a}  + J_{r_1 r_2 }^{3,n} } )} \bigr)} } \Bigr)  
\end{split}
\end{equation}
with $\phi_0^\dag(\tau,\vp)=\phi_0\pdag(\tau,-\vp)$, etc. There are several ways of decoupling the interaction using HS fields; our choice is such that, if the nodal fermions are ignored in \eqref{Action with HS fields}, a saddle-point analysis reproduces the correct Hartree-Fock equations for the antinodal fermions when spin rotation invariance is broken; see \cite{LangmannWallin2:1997} for further discussion of this point. Integrating out the antinodal Grassman fields in \eqref{Action with HS fields} gives a term $-\text{Tr}\ln G^{-1}$, where 
\begin{equation}
\begin{split}
G_{\vk,r,\alpha;\vk',r',\alpha'}^{-1} = &\left( {\partial _\tau   + \varepsilon (\vQ_{r,s}  + \vk) - \mu } \right)\delta_{\vk,\vk'}\delta_{r,r'}\delta_{\alpha,\alpha'} - \frac{U}{2}\left( {\frac{a}{L}} \right)^2
\\
&\times \left( {\ii\hat \phi _0\pdag (\tau ,\vQ_{r',0}  - \vQ_{r,0}  + \vk' - \vk) + \hat \phi_S\pdag (\tau ,\vQ_{r',0}  - \vQ_{r,0}  + \vk' - \vk)\sigma _{\alpha \alpha '}^3 } \right)
\end{split}.
\end{equation}
If we expand this term to quadratic order in the HS fields, we can integrate out these fields and obtain an effective action of nodal fermions. This action can then be analyzed using the same partial continuum limit as in Section~\ref{Sec:Partial continuum limit}.  

\section{\label{Sect: Discussion}Discussion}

In this paper, we have related an effective QFT model of interacting electrons to the 2D Hubbard model near half-filling. The model consists of so-called nodal- and antinodal fermions and is obtained by performing a certain partial continuum limit in the lattice model. We have shown that the nodal part can be studied using bosonization methods in the Hamiltonian framework. Important results are a formula expressing the nodal fermion field operator in terms of Klein factors and density operators, and a 2D extension of the Sugawara construction. We identified a QFT model of 2D electrons (defined by the Hamiltonian in \eqref{Spinfull Mattis Hamiltonian: main}) that can be solved exactly by bosonization. We also obtained a 2D analogue of a Wess--Zumino--Witten model, which we believe is simpler to analyse than the corresponding one in 1D due to different scaling behavior. 

The antinodal fermions can be studied on different levels of sophistication. As in \cite{deWoulLangmann1:2010}, we can do a local-time approximation in the induced antinodal action obtained by integrating out the bosonized nodal fermions. The antinodal fermions can then be studied using ordinary mean field theory. Due to the close similarity to the spinless case \cite{deWoulLangmann1:2010}, we are likely to find a mean field phase near half-filling in which the antinodal fermions are gapped and have an antiferromagnetic ordering. In this partially gapped phase, the low-energy physics of the effective QFT model would then be governed by the nodal fermions alone. 

If the antinodal fermions are gapless, they will contribute to the low-energy physics. As we have proposed above, a crude way to incorporate their effect is to apply a Hubbard-Stratonovich (HS) transformation and then expand the resulting action in powers of the HS fields. This allows us to derive an effective nodal action that becomes Gaussian after bosonization. The study of this action is left to future work.    

\bigskip

\subsection*{Acknowledgments} 
We thank Pedram Hekmati, Vieri Mastropietro, Manfred Salmhofer, Chris Varney and Mats Wallin for useful discussions. This work was supported by the G\"oran Gustafsson Foundation and the Swedish Research Council (VR) under contract no. 621-2010-3708.

\appendix

\section{\label{App: Index sets}Index sets}
The following index sets are used throughout the paper and are collected here for easy reference ($s=\pm$)
\begin{align}
\label{Lattice}
&\Lambda=\{\vx\in a\Z^2 \, : \, -L/2\leq x_\pm<L/2\} \\
\label{Nodal position space}
&\Lambda_s = \left\{\vx\in\R^2\, : \, x_s\in\mathbb{R},\;  x_{-s}\in \ta\mathbb{Z},\; -{L}/{2}\leq x_\pm < {L}/{2} \right\} \\
\label{1D lattice}
&\Lambda_{1\rm D} = \left\{ x \in \ta\mathbb{Z}\; : \;  -{L}/{2}\leq x <{L}/{2} \right\} \\
&\Lambda^*= \{\vk\in \R^2 \, : \, k_\pm\in(2\pi/L)\left(\Z+1/2\right)\} \\
\label{Continuum nodal momentum region}
&\Lambda_{s}^* = \left\{\vk\in \Lambda^* \; : \; -{\pi}/{\ta} \leq k_{-s}< {\pi}/{\ta} \right\} \\
\label{Antinodal momentum region}
&\Lambda_{0}^*= \left\{\vk\in \Lambda^* \, : \, \left|k_\pm + {\pi}/{L} \right| < {\kappa\pi}/({\sqrt{2}a})\right\}\\
&\tilde\Lambda^*= \{\vp\in \R^2 \, : \, p_\pm\in(2\pi/L)\Z\} \\
\label{Boson momentum set}
&\tilde\Lambda_{s}^* = \left\{\vp\in \tilde\Lambda^* \; : \; -{\pi}/{\ta} \leq p_{-s}< {\pi}/{\ta} \right\}\\
\label{Momentum set with zero-component}
&\hat\Lambda_s^*=\left\{\vp\in \tilde\Lambda_s^*\; :\; p_s\neq 0\right\}\\
\label{1D Boson Fourier space}
&\tilde\Lambda_{1\rm D}^* = \left\{ p \in (2\pi/L)\mathbb{Z}\, : \,  -{\pi}/{\ta} \leq p< {\pi}/{\ta} \right\} \\
\label{1D Boson Fourier space with no zero}
&\hat\Lambda_{1\rm D}^* = \left\{ p \in \tilde\Lambda_{1\rm D}^*\, : \,  p \neq 0 \right\}
\end{align}

\section{\label{App:Derivation of the effective model}Derivation of the effective QFT model}

We summarise technical details for the derivation of the nodal-antinodal model from the 2D Hubbard model near half-filling; see also \cite{Langmann2:2010,deWoulLangmann1:2010} for further explanations. 

\subsection{The extended Hubbard model}
\label{App: The extended Hubbard model}
To emphasize the generality of the derivation, we will in this and the following appendices include a nearest-neighbor interaction in the lattice Hamiltonian. We thus consider an extended Hubbard model of itinerant spin-$1/2$ fermions on a diagonal square lattice $\Lambda$ with lattice constant $a$ and $(L/a)^2$ lattice sites (see \eqref{Lattice}). The model is defined by fermion creation- and annihilation operators $\psi_\alpha^{(\dag)}(\vx)$, with $\vx\in\Lambda$ and spin $\alpha=\pm$, acting on a fermion Fock space with vacuum $|0\rangle$ and $\psi\pdag_{\alpha}(\vx)|0\rangle=0$. The fermion operators satisfy antiperiodic boundary conditions and are normalized such that ${\{\psi_\alpha\pdag(\vx),\psi_{\alpha'}^\dag(\vy)\} = \delta_{\alpha ,\alpha'}\delta_{\vx,\vy}/a^2}$, etc. The Hamiltonian is
\begin{equation}
\Hubb \define \sum\limits_{\alpha=\pm}\sum\limits_{\vx,\vy\in\Lambda}a^4\left(-T(\vx-\vy)-\mu\delta_{\vx,\vy}/a^2\right)\psi_\alpha^\dag(\vx) \psi_{\alpha}\pdag(\vy) + \sum\limits_{\vx,\vy\in\Lambda}a^4 u(\vx-\vy) \rho(\vx) \rho(\vy)
\end{equation}
with non-zero hopping matrix elements $T(\vx-\vy)$ equal to $t/a^2>0$ for nn sites and $t'/a^2$ for nnn sites, and non-zero interaction matrix elements $u(\vx-\vy)$ equal to $U/2$ for on-site and $V/2$ for nn sites; the (local) density operators are $\rho(\vx) \define \sum_{\alpha}\psi_\alpha^\dag(\vx) \psi_{\alpha}\pdag(\vx)$. We will assume that the Hubbard parameters satisfy the constraints
\begin{equation}
\label{Hubbard parameter constraints}
|t'|\leq t/2, \qquad U\geq 4V\geq 0.
\end{equation}

We define Fourier-transformed fermion creation- and annihilation operators by
\begin{equation}
\hat\psi_\alpha(\vk) \define \frac{1}{{2\pi}}\sum\limits_{\vx\in\Lambda }{a^2 \psi_\alpha(\vx)\ee^{ - \ii \vk \cdot \vx} } \qquad (\vk\in\Lambda^*)
\end{equation}
such that $\{\hat\psi\pdag_{\alpha}(\vk),\hat\psi^{\dag}_{\alpha'}(\vk')\} = [L/(2\pi)]^2\delta_{\vk,\vk'}\delta_{\alpha,\alpha'}$, etc. Note that the fermion operators in Section~\ref{Sec:Partial continuum limit} are related to these as $\hat c_{\alpha}(\vk)=(2\pi/L)\hat\psi_\alpha(\vk)$. The Fourier-transformed density operators are
\begin{equation}
\begin{split}
\label{Density operators}
\hat\rho(\vp) \define \sum\limits_{\vx\in\Lambda}{a^2\rho(\vx)\ee^{ -\ii \vp\cdot\vx}} 
=\sum\limits_{\alpha=\uparrow,\downarrow}\sum\limits_{\vk_1\vk_2\in\BZ}\sum_{\vn\in\Z^2}\tPiL^2 \hat\psi_{\alpha}^\dag(\vk_1) \hat\psi_{\alpha}\pdag(\vk_2) \delta_{\vk_1+\vp+2\pi\vn/a,\vk_2} 
\end{split}
\end{equation}
where $\vp\in\tilde\Lambda^*$; the last sum in \eqref{Density operators} accounts for umklapp terms and
\begin{equation}
\label{Brillouin zone}
\BZ\define\{\vk\in \Lambda^* \, : \, -\pi/a\leq k_{1,2}<\pi/a\}
\end{equation}
is the Brillouin zone. The Hamiltonian is written in terms of these latter operators as (the second sum is such that $\vp\in(2\pi/L)\Z^2$, $-\pi/a\leq p_{1,2}<\pi/a$)
\begin{equation}
\label{Hubbard Hamiltonian in Fourier space}
\Hubb = \sum\limits_{\alpha=\pm}\sum\limits_{\vk\in\BZ}\tPiL^2\left(\epsilon(\vk)-\mu\right) \hat\psi_{\alpha}^\dag(\vk) \hat\psi_{\alpha}\pdag(\vk)
+\sum\limits_{\vp}\tPiL^2 \hat u(\vp)\hat\rho({-\vp})\hat\rho(\vp)
\end{equation}
with the tight-binding band relation in \eqref{Tight-binding band relation} and
\begin{equation}
\label{Hubbard interaction vertex}
\hat u(\vp) = a^2\Bigl(U/2 + V\bigl[\cos \left( ap_1\right) + \cos\left( ap_2\right) \bigr]\Bigr)/\left( {2\pi}\right)^2
\end{equation} 
the interaction potential. With the chosen normalization for $\hat\psi^{(\dag)}_{\alpha}(\vk)$, the fermion number operator is given by
\begin{equation}
\label{Number operator}
N=\hat\rho(\vzero) =\sum\limits_{\alpha=\pm}\sum\limits_{\vk\in\BZ}\tPiL^2\hat\psi_{\alpha}^\dag(\vk) \hat\psi_{\alpha}\pdag(\vk).
\end{equation}
We recall that, under a particle-hole transformation
\begin{equation}
\label{Particle-hole transformation}
\cW_{ph}\pdag \hat \psi_\alpha\pdag(\vk)\cW_{ph}^\dag \define \hat\psi_\alpha ^\dag\left(-\vk + \left(\pi,\pi\right)/a\right),
\end{equation}
the filling $\nu$ is mapped to $2-\nu$, while the Hamiltonian in \eqref{Hubbard Hamiltonian in Fourier space} transforms as
\begin{equation}
\begin{split}
\cW_{ph}\pdag \Hubb{\left(t,t',\mu,U,V\right)} \cW_{ph}^\dag   = \Hubb{\left(t, - t',2(U + 4V) - \mu ,U,V\right)} 
\\
+ 2\left( U + 4V - \mu\right)\left( L/a \right)^2 
\end{split}
\end{equation}
where the notation $\Hubb{\left(t,t',\mu,U,V\right)}$ has been used for the right hand side of \eqref{Hubbard Hamiltonian in Fourier space}.

\subsection{\label{App:Eight-flavor representation of the Hubbard Hamiltonian}Eight-flavor representation of the Hamiltonian}
We rewrite the Hubbard Hamiltonian in terms of nodal-, antinodal-, in- and out fermions. To this end, we let $\mathcal{I}$ be the index set of the eight pairs of flavor indices $(r,s)$, with $r=\pm$ and $s=0,\pm,2$. The momentum regions are defined as ($r=\pm$)
\begin{equation}
\label{Momentum sets for different flavors}
\begin{split}
&\Lambda_{r,0}^*\define \left\{\vk\in \Lambda^* \, : \, \left|k_\pm + {\pi}/{L} \right| < {\kappa\pi}/({\sqrt{2}a})\right\}
\\
&\Lambda_{r,2}^*\define \left\{\vk\in \Lambda^* \, : \, \left|k_\pm + {\pi}/{L} \right| < {\pi}/{\ta} \right\}
\\
& \Lambda_{r,s=\pm}^*\define \left\{\vk\in \Lambda^* \, : \, \left|k_s + \frac{\pi}{L} + r\frac{2Q-\pi}{\sqrt{2}a} \right| < \frac{\kappa\pi}{\sqrt{2}a}, \, \left|k_{-s} + \frac{\pi}{L}\right| < \frac{\pi}{\ta}\right\}
\end{split}
\end{equation}
with the parameters 
\begin{equation}
\label{kappa, Q and atilde}
\kappa\in(2\sqrt{2}a/L)(\N+1/2), \qquad Q\in(\sqrt{2}\pi a/L)\N, \qquad \ta \define \sqrt{2}a/(1-\kappa)
\end{equation}
satisfying the geometric constraints
\begin{equation}
\label{Geometric constraints}
Q\neq\pi/2, \qquad \pi(1-\kappa)/2<Q<\pi(1+\kappa)/2, \qquad 0<\kappa < 1.
\end{equation}
The relative number of momenta in these regions, defined as $f_{r,s}\define\sum_{\vk\in\Lambda^*_{r,s}}(a/L)^2$, are 
\begin{equation}
f_{r,0}=\kappa^2/2,\qquad f_{r,2}=(1-\kappa)^2/2,\qquad f_{r,s=\pm}=\kappa(1-\kappa)/2.
\end{equation}
We also set
\begin{gather}
\begin{aligned}
&\vQ_{-,2}\define(0,0),& \quad &\vQ_{+,2}\define\left(\pi/a,\pi/a\right),& \quad & \vQ_{r,s=\pm}\define\left(rQ/a,rsQ/a\right) \\ 
&\vQ_{+,0}\define\left(\pi/a,0\right),& \quad &\vQ_{-,0}\define\left(0,\pi/a\right),& \quad & \phantom{-}
\end{aligned}
\end{gather} 
and define new fermion operators by 
\begin{equation}
\hat\psi_{r,s,\alpha}^{(\dag)}(\vk)\define \hat\psi_{\alpha}^{(\dag)}(\vQ_{r,s} + 2\pi\vn/a + \vk) \qquad (\vk\in\Lambda_{r,s}^*)
\end{equation}
with $\vn\in\Z^2$ such that $\vQ_{r,s} + 2\pi\vn/a + \vk\in\BZ$. They satisfy the anticommutation relations 
\begin{equation}
\label{Fermion flavor anticommutation relations}
\{\hat\psi\pdag_{r,s,\alpha}(\vk),\hat\psi^{\dag}_{r',s',\alpha'}(\vk')\} = [L/(2\pi)]^2\delta_{r,r'}\delta_{s,s'}\delta_{\alpha,\alpha'}\delta_{\vk,\vk'}, \quad \{\hat\psi\pdag_{r,s,\alpha}(\vk),\hat\psi\pdag_{r',s',\alpha'}(\vk')\} = 0.
\end{equation}
In terms of these operators, the kinetic part of the Hubbard Hamiltonian in \eqref{Hubbard Hamiltonian in Fourier space} can be written
\begin{equation}
\label{Kinetic part in eight-flavor representation}
\HubbKin=\sum\limits_{\alpha=\pm}\sum\limits_{(r,s)\in\mathcal{I}}\sum\limits_{\vk\in\Lambda^*_{r,s}}\tPiL^2\left[\epsilon(\vQ_{r,s}+\vk)-\mu\right]\hat\psi_{r,s,\alpha}^{\dag}(\vk)\hat\psi_{r,s,\alpha}\pdag(\vk)
\end{equation}
and similarly for the interaction part
\begin{equation}
\label{Interaction part in eight-flavor representation}
\begin{split}
\HubbInt=\sum\limits_{\alpha,\alpha'=\pm}\sum\limits_{(r_j,s_j)\in\mathcal{I}}\sum\limits_{\vk_j\in\Lambda^*_{r_j,s_j}}&\tPiL^8\hat v(K_1,K_2,K_3,K_4)
\\
&\times\hat\psi_{r_1,s_1,\alpha}^{\dag}(\vk_1)\hat\psi_{r_2,s_2,\alpha}\pdag(\vk_2)\hat\psi_{r_3,s_3,\alpha'}^{\dag}(\vk_3)\hat\psi_{r_4,s_4,\alpha'}\pdag(\vk_4)
\end{split}
\end{equation}
with $K_j$ short for $\vQ_{r_j,s_j}+\vk_j$, and
\begin{equation}
\begin{split}
\label{Interaction vertex in eight-flavor representation} 
\hat v(K_1,K_2,K_3,K_4) = &\hat u(\vQ_{r_1,s_1}-\vQ_{r_2,s_2}+\vk_1-\vk_2)
\\
&\times\sum\limits_{\vn\in\Z^2}\LtPi^2 \delta_{\vQ_{r_1,s_1}-\vQ_{r_2,s_2}+\vQ_{r_3,s_3}-\vQ_{r_4,s_4}+\vk_1-\vk_2+\vk_3-\vk_4,2\pi\vn/a}
\end{split}.
\end{equation}
Setting $V=0$ gives \eqref{Kinetic part in eight-flavor representation: main}--\eqref{Interaction part in eight-flavor representation: main}. The fermion number operator can be expressed as
\begin{equation}
N= \sum\limits_{(r,s)\in\mathcal{I}} N_{r,s}, \qquad N_{r,s} \define \sum\limits_{\alpha=\pm}\sum\limits_{\vk\in\Lambda^*_{r,s}}\tPiL^2\hat\psi_{r,s,\alpha}^{\dag}(\vk)\hat\psi_{r,s,\alpha}\pdag(\vk).
\end{equation}
Note that the mapping from the representation in \eqref{Hubbard Hamiltonian in Fourier space} to the one in \eqref{Kinetic part in eight-flavor representation}--\eqref{Interaction vertex in eight-flavor representation} is exact.

Under the particle-hole transformation defined in \eqref{Particle-hole transformation}, the fermion operators in the eight-flavor representation transform as 
\begin{equation}
\cW_{ph}\pdag \hat \psi_{r,0,\alpha}\pdag(\vk)\cW_{ph}^\dag  = \hat \psi_{-r,0,\alpha}^\dag(-\vk), \qquad \cW_{ph}\pdag \hat \psi_{r,2,\alpha}\pdag(\vk)\cW_{ph}^\dag  = \hat \psi_{-r,2,\alpha}^\dag(-\vk)
\end{equation}
and
\begin{equation}
\cW_{ph}\pdag \hat \psi_{r,s,\alpha}\pdag(\vk)\cW_{ph}^\dag  = \hat \psi_{r,s,\alpha}^\dag( -\vk+\left(\pi,\pi\right)/a - 2{\vQ_{r,s}} + 2\pi \vn/a), \qquad (s=\pm)
\end{equation}
where $\vn\in\Z^2$ is such that $(-\vk+\left(\pi,\pi\right)/a - 2{\vQ_{r,s}} + 2\pi \vn/a)\in\Lambda_{r,s}^*$.

\subsection{\label{App:Simplified matrix elements}Simplified matrix elements}
Expanding the tight-binding band relation to lowest non-trivial order around the points $\vQ_{r,s}$, $(r,s)\in\mathcal{I}$, leads to
\begin{equation}
\epsilon(\vQ_{r,s} + \vk) = \epsilon(\vQ_{r,s}) + \varepsilon_{r,s}(\vk) + \ldots \qquad (\vk\in\Lambda_{r,s}^*)
\end{equation}
with constants
\begin{equation}
\label{Band relation constants}
\epsilon(\vQ_{r,0})=4t', \qquad \epsilon(\vQ_{r,\pm}) =  -4\cos(Q)\left[{t + t'\cos (Q)}\right], \qquad \epsilon(\vQ_{r,2})=4(rt-t')
\end{equation}
and effective band relations
\begin{equation}
\label{Effective band relations}
\varepsilon_{r,0}(\vk)= -rc_Fk_+k_- - c_F'|\vk|^2,\quad \varepsilon_{r,\pm}(\vk)=rv_Fk_\pm,\quad \varepsilon_{r,2}(\vk)=\left(-rc_F/2+c_F'\right)|\vk|^2
\end{equation}
where
\begin{equation}
\label{Effective band relations: constants}
v_F\define2\sqrt{2}\sin(Q)\left[t+2t'\cos(Q)\right]a, \qquad c_F \define 2ta^2, \qquad c_F'\define2t'a^2.
\end{equation}

Thus, the nodal fermions have (approximately) a linear-, the antinodal fermions a hyperbolic-, and the in- and out fermions a parabolic band relation. 

\begin{approximation}
\label{Approx: Band relation}
Replace the tight-binding band relations $\epsilon(\vQ_{r,s} + \vk)$ in \eqref{Kinetic part in eight-flavor representation} by $\epsilon(\vQ_{r,s}) + \varepsilon_{r,s}(\vk)$ defined in Equations \eqref{Band relation constants}--\eqref{Effective band relations: constants}.
\end{approximation}

We note that this approximation is only crucial for the nodal fermions and is not done for the other fermions in the main text.  

\begin{approximation}
\label{Approx: Vertex}
Simplify the interaction vertex in \eqref{Interaction vertex in eight-flavor representation} by replacing the right-hand side with
\begin{equation}
\label{Truncated interaction vertex in eight-flavor representation} 
\hat u(\vQ_{r_1,s_1}-\vQ_{r_2,s_2})\LtPi^2\delta_{\vk_1-\vk_2+\vk_3-\vk_4,\vzero}\sum\limits_{\vn\in\Z^2}\delta_{\vQ_{r_1,s_1}-\vQ_{r_2,s_2}+\vQ_{r_3,s_3}-\vQ_{r_4,s_4},2\pi\vn/a}.
\end{equation}
\end{approximation}

In addition to the added constraint \eqref{Q-vector constraint}, this involves expanding the interaction vertex in \eqref{Interaction vertex in eight-flavor representation} as
\begin{equation}
\label{Interaction vertex expansion}
\hat{u}(\vQ_{r,s} -\vQ_{r',s'} + \vk-\vk') = \hat{u}(\vQ_{r,s} -\vQ_{r',s'}) + O\left(|a(\vk-\vk')|\right)
\end{equation}
and then only keeping the lowest-order term; this approximation is not needed if $V=0$. Again, Approximation~\ref{Approx: Vertex} is only crucial for scattering processes involving nodal fermions and will not be done for processes involving only antinodal fermions in the main text. 

The constraint imposed by the second Kronecker delta in \eqref{Truncated interaction vertex in eight-flavor representation} reduces the number of terms in the original Hubbard interaction: of the originally 4096 possible combinations of pairs $(r_j,s_j)$, 512 yield a non-zero interaction vertex if $Q=\pi/2$, and $196$ if $Q\neq\pi/2$. The combinations of $(r_j,s_j)\in\mathcal{I}$ for which \eqref{Truncated interaction vertex in eight-flavor representation} is non-zero when $Q\neq\pi/2$ are collected in Table~\ref{Flavor combinations}.

\begin{table}[ht!]
\vspace{0.1cm}
  \begin{center}
    \begin{tabular}{|c|c|c|c||c||c|}
      \hline & & & & &
      \\[-1.0ex] 
      $r_1,s_1$ & $r_2,s_2$ & $r_3,s_3$& $r_4,s_4$ & Restrictions & \#
      \\[1.2ex]
      \hline & & & & &
      \\[-1.0ex]
      $r,s$ & $r,s$ & $r,s$ & $r,s$ & $s =0,\pm,2$ & \phantom{0}8    
      \\      
			$r,s$ & $r,s$ & $r',s'$ & $r',s'$ & $(r,s)\neq (r',s')$, $s,s'=0,\pm,2$ & 56
      \\
      $r,s$ & $r',s'$ & $r',s'$ & $r,s$ & $(r,s)\neq (r',s')$, $s,s'=0,\pm,2$ & 56
      \\      
      $r,s$ & $r',s'$ & $-r,s$ & $-r',s'$ & $(s,s')=(\pm,\mp), (0,2), (2,0)$ & 16
      \\
      $r,s$ & $-r,s$ & $r,s$ & $-r,s$ & $s=0,2$ & \phantom{0}4
      \\
      $r,s$ & $r',s'$ & $r,s$ & $-r',s'$ & $s=0,2$, $s'=\pm$ & 16
      \\
      $r,s$ & $r',s'$ & $-r,s$ & $r',s'$ & $s=\pm$, $s'=0,2$ & 16
      \\
      $r,s$ & $r',s'$ & $r,s$ & $r',s'$ & $(s,s')=(0,2), (2,0)$ & \phantom{0}8
      \\
      $r,s$ & $r',s'$ & $-r',s'$ & $-r,s$ & $(s,s')=(0,2), (2,0)$ & \phantom{0}8
      \\
      $r,s$ & $-r,s$ & $r',s'$ & $-r',s'$ & $(s,s')=(0,2), (2,0)$ & \phantom{0}8
      \\ 
[1.2ex] 
      \hline
    \end{tabular}
  \end{center}  
\caption{\label{Flavor combinations}List of all combinations for $(r_j,s_j)\in\mathcal{I}$ that satisfy the constraint in \eqref{Q-vector constraint} when $Q\neq \pi/2$; $r,r'=\pm$. The rightmost column indicates the total number of terms corresponding to each line; they sum up to 196.}
\end{table}

Define interaction coefficients $v_{r,s,r',s'}\define a^2\hat u(\vQ_{r,s}-\vQ_{r',s'})/(2\pi)^2$, with
\begin{gather}
\begin{aligned}
&v_{+,0,-,0} = v_{+,2,-,2} =U/2 - 2V,& \quad &v_{+,\pm,-,\pm} = U/2 + 2V\cos\left(2Q\right) \\ 
&v_{r,0,r',\pm} = v_{r,0,r',2} = U/2,& \quad &v_{r,\pm ,r',\mp} = U/2 + V\left(1 + \cos \left(2Q\right)\right)\\
&v_{r,\pm,r',2} = U/2 - r'2V\cos\left(Q \right),& \quad &v_{r,s,r,s} = U/2 + 2V
\end{aligned}
\end{gather} 
for $r,r'=\pm$. Introducing Approximations~\ref{Approx: Band relation} and \ref{Approx: Vertex} into Equations~\eqref{Kinetic part in eight-flavor representation}--\eqref{Interaction vertex in eight-flavor representation} leads to the truncated Hamiltonian 
\begin{equation}
\TruncHubb = \TruncHubbKin + \TruncHubbInt
\end{equation}
with
\begin{equation}
\TruncHubbKin = \sum\limits_{\alpha=\pm}\sum\limits_{(r,s)\in\mathcal{I}}\sum\limits_{\vk\in\Lambda^*_{r,s}}\tPiL^2\left(\varepsilon_{r,s}(\vk)-\left[\mu-\epsilon(\vQ_{r,s})\right]\right)\hat\psi_{r,s,\alpha}^{\dag}(\vk)\hat\psi_{r,s,\alpha}\pdag(\vk)
\end{equation}
and
\begin{equation}
\label{Truncated Hubbard interaction}
\begin{split}
\TruncHubbInt = \sum\limits_{\genfrac{}{}{0pt}{1}{(r,s),(r',s')\in\mathcal{I}}{(r,s)\neq(r',s')}}f_{r,s}v_{r,s,r',s'}N_{r',s'}+\sum\limits_{(r,s),(r',s')\in\mathcal{I}} \sum\limits_{\vp\in \tilde\Lambda^*}  \frac{1}{L^2} \Bigl( g_{r,s,r',s'}^{\Charge}{\hat \rho}_{r,s}^\dag{\hat \rho}_{r',s'}\pdag 
\\
+ g_{r,s,r',s'}^{\SpinZ}{\hat {\bf S}}_{r,s}^\dag\cdot{\hat {\bf S}}_{r',s'}\pdag + g_{r,s,r',s'}^{\Pair}{\hat \Pair}_{r,s}^\dag\cdot\hat{\Pair}\pdag_{r',s'} \Bigr) + \tilde H_{rem}
\end{split}
\end{equation}
where $\tilde H_{rem}$ contains interaction terms between in- or out fermions and nodal- or antinodal fermions (including e.g.\ the last three lines in Table~\ref{Flavor combinations}),
\begin{align}
\label{Flavour density operators}
\hat \rho_{r,s}\pdag(\vp) & \define  \sum\limits_{\alpha=\pm}\sum\limits_{\vk_1,\vk_2\in\Lambda^*_{r,s}} \tPiL^2 \hat\psi^\dag_{r,s,\alpha}(\vk_1)\hat\psi\pdag_{r,s,\alpha}(\vk_2)\delta_{\vk_1+\vp,\vk_2}
\\
\label{Flavour spin operators}
\hat S_{r,s}^i(\vp) & \define  \frac12\sum\limits_{\alpha,\beta=\pm}\sum\limits_{\vk_1,\vk_2\in\Lambda^*_{r,s}} \tPiL^2 \hat\psi^\dag_{r,s,\alpha}(\vk_1)\sigma_{\alpha,\beta}^i\hat\psi\pdag_{r,s,\beta}(\vk_2)\delta_{\vk_1+\vp,\vk_2} \end{align} 
such that ${\hat {\bf S}}_{r,s}^\dag\cdot{\hat {\bf S}}_{r',s'}\pdag = \sum_{i=1}^3 \hat S_{r,s}^i(-\vp) \hat S_{r',s'}^i(\vp)$, and
\begin{equation}
\label{Flavour pair operators}
\hat \Pair_{r,s}^\mu(\vp) \define  \frac12\sum\limits_{\alpha,\beta=\pm}\sum\limits_{\vk_1\in\Lambda^*_{r,s}}\sum\limits_{\vk_2\in\Lambda^*_{r_s,s}} \tPiL^2 \hat\psi\pdag_{r_s,s,\alpha}(\vk_1)\sigma_{\alpha,\beta}^\mu\hat\psi\pdag_{r,s,\beta}(\vk_2)\delta_{\vk_1+\vk_2,\vp}
\end{equation} 
where $r_s\equiv r$ for $s=0,2$ (antinodal-, in-, and out fermions), $r_s\equiv -r$ for $s=\pm$ (nodal fermions), and ${\hat \Pair}_{r,s}^\dag\cdot\hat{\Pair}\pdag_{r',s'} = \sum_{\mu=0}^3 [\Pair_{r,s}^\mu(\vp)]^\dag \Pair_{r',s'}^\mu(\vp)$. The coupling constants are
\begin{equation}
\label{Coupling constants for effective Hamiltonian}
\begin{split}
g_{r,s,r',s'}^\Charge  \define & \, a^2\left[\delta_{r,r'}\delta_{s,s'}v_{r,s,r',s'} + \left(1 - \delta_{r,r'}\delta _{s,s'}\right)\left(v_{r,s,r,s} - v_{r,s,r',s'}/2 \right)\right]
\\
g_{r,s,r',s'}^\SpinZ  \define & \,  2a^2\left(\delta _{r,r'}\delta _{s,s'}-1\right)v_{r,s,r',s'}
\\
g_{r,s,r',s'}^\Pair  \define & \, 2a^2 \Bigl[ {\delta _{s, - s'} \left( {\delta _{s, + }  + \delta _{s, - } } \right) + \delta _{r, - r'} \delta _{s,s'} \left( {\delta _{s,0}  + \delta _{s,2} } \right)}\\ 
&\qquad{ + \left( {\delta _{s, + }  + \delta _{s, - } } \right)\delta _{s',0}  + \delta _{s,0} \left( {\delta _{s', + }  + \delta _{s', - } } \right)}\Bigr]v_{r,s,r',s'}
\end{split}.
\end{equation}

\subsection{\label{App:Normal-ordering}Normal-ordering}
Depending on the filling of the system, different degrees of freedom will be important for the low-energy physics. To distinguish these, we define a reference state (Fermi sea)
\begin{equation}
\label{Fermi sea}
\Omega \define \prod_{\alpha=\pm}\prod_{\vk\in\cS}\psi_{\alpha}^\dag(\vk) |0\rangle
\end{equation}
with the set $\cS\subset\BZ$ chosen such that one of the following three cases hold: (I) all nodal-, antinodal- and out fermion states are unoccupied with the filling $\nu\ll 1$, (II) all in-, nodal- and antinodal fermion states are occupied with $\nu\gg 1$, or (III) all in fermion states are occupied and all out fermion states are unoccupied with $\nu\approx 1$. 

The filling factors of the state \eqref{Fermi sea} for different fermion flavors are defined as $\nu_{r,s}\define (a/L)^2\langle \Omega, N_{r,s}\Omega\rangle$ with the total filling $\nu=\sum_{(r,s)\in\mathcal{I}}\nu_{r,s}$. 
By fermion normal-ordering the bilinears in \eqref{Flavour density operators}--\eqref{Flavour pair operators} with respect to \eqref{Fermi sea}, one finds ($i=1,2,3$ and $\mu=0,1,2,3$)
\begin{equation}
\label{Normal-ordered density and spin operators}
\hat J_{r,s}^0\define \,:\! \hat \rho_{r,s}\!:\, = \hat \rho_{r,s} - (L/a)^2\nu_{r,s}\delta_{\vp,\vzero}, \qquad :\! \hat S^i_{r,s}\!:\,=\hat S^i_{r,s},\qquad :\! \hat \Pair^\mu_{r,s}\!:\,=\hat \Pair^\mu_{r,s}
\end{equation}
where $\hat J_{r,s}^0=\hat J_{r,s}^0(\vp)$, etc. We note that the terms in $\tilde H_{rem}$ are automatically normal-ordered.

\begin{approximation}
Drop all normal-ordered interaction terms between in- and out fermions, and between in- or out fermions and nodal- or antinodal fermions.
\end{approximation}

This approximation leads to the following (eight-flavor) Hamiltonian consisting of decoupled in-, out-, and nodal-antinodal fermions
\begin{equation}
\label{Effective Hamiltonian in the eight-flavor representation}
\begin{split}
&\HEightFlavor\define \cE + \sum\limits_{\alpha=\pm}\sum\limits_{r,s\in\cI}\sum\limits_{\vk\in\Lambda^*_{r,s}}\tPiL^2\left[\varepsilon_{r,s}(\vk)-\mu_{r,s}\right] :\! \hat\psi^\dag_{r,s,\alpha}(\vk) \hat\psi\pdag_{r,s,\alpha}(\vk)\!:
\\ 
& + \sum\limits_{r,r',s,s'} \sum\limits_{\vp\in\tilde\Lambda^*} \frac{1}{L^2}\Bigl( g_{r,s,r',s'}^{\Charge}{\hat J}_{r,s}^{0\dag}{\hat J}_{r',s'}^0 -g_{r,s,r',s'}^{\SpinZ}{\hat {\bf S}}_{r,s}^\dag\cdot{\hat {\bf S}}_{r',s'}\pdag + g_{r,s,r',s'}^{\Pair}{\hat \Pair}_{r,s}^\dag\cdot\hat{\Pair}\pdag_{r',s'} \Bigr)
\\
& + \sum\limits_{r} \sum\limits_{\vp\in\tilde\Lambda^*} \frac{1}{L^2}
\Bigl( g_{r,2,r,2}^{\Charge}{\hat J}_{r,2}^{0\dag}{\hat J}_{r,2}^0 -g_{r,2,r,2}^{\SpinZ}{\hat {\bf S}}_{r,2}^\dag\cdot{\hat {\bf S}}_{r,2}\pdag + g_{r,2,r,2}^{\Pair}{\hat \Pair}_{r,2}^\dag\cdot\hat{\Pair}\pdag_{r,2} \Bigr)
\end{split}
\end{equation}
with the sums in the second line such that $s,s'=0,\pm$ (the nodal- and antinodal interaction terms),
\begin{equation}
\label{Effective chemical potential}
\mu_{r,s} =  \mu - \varepsilon (Q_{r,s}) - \sum\limits_{\genfrac{}{}{0pt}{1}{(r',s')\in\mathcal{I}}{(r',s')\neq(r,s)}} {\left( {f_{r',s'}  - \nu _{r',s'} } \right)v_{r,s,r',s'} }  - 2\nu v_{r,s,r,s}  
\end{equation}
the effective chemical potentials, and
\begin{equation}
\label{Energy constant in effective Hamiltonian}
\begin{split}
&\cE = \sum\limits_{\alpha=\pm} \sum\limits_{(r,s)\in\mathcal{I}} \sum\limits_{\vk\in\Lambda^*_{r,s}} \tPiL^2 \left[\epsilon(\vQ_{r,s})+\varepsilon_{r,s}(\vk)\right] \langle\Omega,\hat\psi_{r,s,\alpha}^{\dag}(\vk)\hat\psi_{r,s,\alpha}\pdag(\vk)\Omega\rangle + \cE_1
\\
&\Bigl(\frac{a}{L}\Bigr)^2\cE_1 \define - \mu \nu + (U/2 + 2V)\nu^2 + \sum\limits_{\genfrac{}{}{0pt}{1}{(r,s),(r',s')\in\mathcal{I}}{(r,s)\neq(r',s')}} {\nu_{r,s} \left( {f_{r',s'}  - \nu _{r',s'} /2} \right)v_{r,s,r',s'} }   
\end{split}
\end{equation}
an additive energy constant. 

\subsection{\label{App: Partial continuum limit in underdoped regime}Partial continuum limit near half-filling}
In this paper, we will concentrate on the nearly half-filled regime for which the in- and out fermions can be ignored in \eqref{Effective Hamiltonian in the eight-flavor representation} (corresponding to case (III) above). To this end, we choose the momentum set $\cS$ in \eqref{Fermi sea} such that
\begin{equation}
\hat\psi_{-,2,\alpha }^\dag(\vk)\Omega = 0 \quad \mbox{for all }\vk\in\Lambda_{-,2}^*, \qquad \hat\psi_{+,2,\alpha }\pdag(\vk)\Omega = 0 \quad \mbox{for all }\vk\in\Lambda_{+,2}^*
\end{equation}
for the in- and out fermions, 
\begin{equation}
\hat\psi\pdag_{r,0}(\vk)\Omega = 0 \quad \mbox{for all } \vk\in\Lambda_{r,0}^*. 
\end{equation}
for the antinodal fermions, and
\begin{equation}
\hat\psi_{r,s,\alpha }^\dag(\vk)\Omega = \hat\psi_{r,s,\alpha}\pdag(-\vk)\Omega=0  \qquad \mbox{for all } \vk\in\Lambda_{r,s}^*\, : \, rk_s \leq \sqrt{2}\left(Q_0-Q\right)/a 
\end{equation}
for the nodal fermions ($s=\pm)$; the parameter $Q_0$ satisfy the same requirements as $Q$ in \eqref{kappa, Q and atilde}--\eqref{Geometric constraints}. With this, the filling factors of \eqref{Fermi sea} become
\begin{equation}
\begin{split}
\nu_{r,0} = 0, \quad \nu _{r,\pm}=\left(1 - \kappa\right)\left({2Q_0}/\pi - 1 + \kappa\right)/2, \quad \nu_{-,2} = \left(1 - \kappa\right)^2, \quad \nu_{+,2} = 0
\end{split}
\end{equation}
such that the total filling is $\nu  = 1 -\kappa^2 + 2\left(1 - \kappa\right)\left({2Q_0}/{\pi}-1 \right)$.

The chemical potential $\mu$ is fixed such that $\varepsilon_{r,s}(\vk)-\mu_{r,s}=0$ for $s=\pm$ and momenta $\vk$ satisfying $\vk+\vQ_{r,s} = (rQ_0/a,rsQ_0/a)$, i.e. 
\begin{equation}
\label{Chemical potential constraint}
v_F\sqrt{2}\left(Q_0-Q\right)/a-\mu_{r,s}=0. 
\end{equation}
This is equivalent to requiring that the underlying Fermi surface corresponding to \eqref{Fermi sea} crosses the points $(rQ_0/a,rsQ_0/a)$. One finds
\begin{equation}
\begin{split}
\mu  = v_F\sqrt{2}\left({Q_0}-Q\right)/a - 4t \cos\left(Q\right) - 4t' {\cos^2}\left(Q\right) + {U}/{2} - 4VC\cos\left(Q\right) 
\\
+ \left({1 - \kappa}\right)\left({{2{Q_0}}/{\pi} - 1}\right)\left({U}/{4} + V\right) + \left({U}/{2} + 4V \right)\nu 
\end{split}
\end{equation}
with
\begin{equation}
C \define \left({1 - \kappa}\right)\cos\left(Q\right)\left({2{Q_0}}/{\pi} - 1 \right) + {\left({1-\kappa}\right)^2}/2.
\end{equation}
Likewise, the energy constant $\cE_1$ in \eqref{Energy constant in effective Hamiltonian} becomes
\begin{equation}
\begin{split}
\Bigl(\frac{a}{L}\Bigr)^2\cE_1 =& -\mu\nu +\left(U + 4V\right)\nu^2/{2}- 4V{C^2} + U{\kappa^2}\left({1-\kappa}\right)\left({{2{Q_0}}/{\pi} - 1}\right)
\\
& + U\left({1 - \kappa}\right)\left({2{\kappa^3}+ \kappa+ 1}\right)/4 + V{\kappa^2}{\left({1-\kappa}\right)^2}\left({4{{\cos}^2}\left(Q \right)- 1}\right)
\\
& + \left({V-{3U}/{4}}\right){\left({1 - \kappa}\right)^2}{\left({{2{Q_0}}/{\pi }- 1}\right)^2}
\end{split}.
\end{equation}
Within a mean field approximation, one can fix the parameter $Q$ using \eqref{Chemical potential constraint} and imposing the self-consistency condition $Q=Q_0$; the reader is referred to \cite{deWoulLangmann1:2010} for details. In the following, we simplify the presentation by taking $Q=Q_0$ at the outset (thus setting $\mu_{r,s=\pm}=0$) at the cost of keeping $Q$ as a free parameter. 

\begin{approximation}
Drop all terms in \eqref{Effective Hamiltonian in the eight-flavor representation} involving in- and out fermions.
\end{approximation}

We regularize the interaction in \eqref{Effective Hamiltonian in the eight-flavor representation} using the cutoff functions (see Section~5.4 in \cite{Langmann2:2010} for further discussion of these functions)
\begin{equation}
\label{General cutoff functions}
\chi_s\pdag(\vp)\define\begin{cases}
1& \text{if } \left|p_s\right|< \kappa\pi/(\sqrt{2}a) \text{ and } \left|p_{-s}\right|<\pi/\ta\\
0& \text{otherwise}
\end{cases} 
\end{equation}
for $s=\pm$ and $\vp\in\tilde\Lambda^*$; we use a somewhat simplified cutoff function in the main text.

\begin{approximation}
Replace all nodal operators ${\hat J}_{r,s}^0(\vp)$, ${\hat S^i}_{r,s}(\vp)$, and ${\hat \Pair}_{r,s}^\mu(\vp)$ ($s=\pm$) in \eqref{Effective Hamiltonian in the eight-flavor representation} by the operators $\chi_s(\vp){\hat J}_{r,s}^0(\vp)$, $\chi_s(\vp){\hat S^i}_{r,s}(\vp)$, and $\chi_s(\vp){\hat \Pair}_{r,s}^\mu(\vp)$.
\end{approximation}

With this, the UV cutoff can be partly removed for the nodal fermions:

\begin{approximation}
Replace the nodal momentum sets $\Lambda_{r,s=\pm}^*$ in \eqref{Momentum sets for different flavors} by $\Lambda_{s=\pm}^*$ in \eqref{Continuum nodal momentum region}.    
\end{approximation}

We will use the same notation for the reference state in \eqref{Fermi sea} defined before taking the partial continuum limit, and the Dirac vacuum obtained after the limit.

In order to facilitate the bosonization of the nodal Hamiltonian (see the discussion in Section~\ref{Sec:Partial continuum limit}), we need to add certain umklapp terms to the nodal density- and spin bilinears. 
\begin{approximation}
Replace the fermion normal-ordered nodal density- and spin operators in \eqref{Flavour density operators}--\eqref{Flavour spin operators} (using \eqref{Normal-ordered density and spin operators}) by 
\begin{align}
\hat J_{r,s}^0(\vp) & \define  \sum\limits_{\alpha}\sum\limits_{\vk_1,\vk_2\in\Lambda^*_{s}} \tPiL^2 :\!\hat\psi^\dag_{r,s,\alpha}(\vk_1)\hat\psi\pdag_{r,s,\alpha}(\vk_2)\!:
\sum\limits_{n\in\Z}\delta_{\vk_1+\vp,\vk_2 + 2\pi n\ve_{ - s}/\ta}
\\
\hat S_{r,s}^i(\vp) & \define  \frac12\sum\limits_{\alpha,\beta}\sum\limits_{\vk_1,\vk_2\in\Lambda^*_{s}} \tPiL^2 :\!\hat\psi^\dag_{r,s,\alpha}(\vk_1)\sigma_{\alpha,\beta}^i\hat\psi\pdag_{r,s,\beta}(\vk_2)\!:
\sum\limits_{n\in\Z}\delta_{\vk_1+\vp,\vk_2 + 2\pi n\ve_{ - s}/\ta}.
\end{align}
\end{approximation}

After applying all this to \eqref{Effective Hamiltonian in the eight-flavor representation}, the effective Hamiltonian of the coupled system of nodal- and antinodal fermions becomes
\begin{equation}
\label{Effective Hamiltonian near half-filling}
\HPartContLim \define \Hn + \Ha + \Hantnod + \cE
\end{equation}
with the nodal part of \eqref{Effective Hamiltonian near half-filling} given by 
\begin{equation}
\begin{split}
\label{Regularised nodal Hamiltonian: app}
&\Hn = \HWZW + g_n^\Pair \sum\limits_{r,r',s=\pm}\sum\limits_{\vp\in\tilde\Lambda^*}\frac{1}{L^2}\chi_+\pdag(\vp)\chi_-\pdag(\vp)\hat\Pair_{r,s}^\dag(\vp)\cdot\hat \Pair_{r',-s}\pdag(\vp)\\
&\HWZW = \HWZWKin+\HWZWInt
\end{split}
\end{equation}
with
\begin{equation}
\label{WZW Hamiltonian: Kinetic term: app}
\HWZWKin = v_F \sum\limits_{\alpha=\pm}\sum\limits_{r,s=\pm} \sum\limits_{\vk\in\Lambda_s^*} \Bigl({\frac{{2\pi }}{L}}\Bigr)^2 rk_s :\hat\psi_{r,s,\alpha }^\dag(\vk)\hat\psi_{r,s,\alpha}\pdag(\vk): 
\end{equation}
the free part, and
\begin{equation}
\begin{split}
\label{WZW Hamiltonian: Interaction term: app}
\HWZWInt = \sum\limits_{\vp\in\tilde\Lambda^*}\frac{1}{L^2} \Bigl( \sum\limits_{s =\pm}\chi_s\pdag(\vp)\bigl(\sum\limits_{r=\pm} g_0^\Charge \hat J_{r,s}{0^\dag} \hat J_{r,s}^0 
+ g_1^\Charge \hat J_{+,s}^{0\dag}\hat J_{-,s}^0+ g_1^\SpinZ {\hat {\bf S}}_{+,s}^\dag \cdot {\hat {\bf S}}_{-,s}\pdag\bigr)
\\
+ \chi_+\pdag(\vp)\chi_-\pdag(\vp)\sum\limits_{r,r'=\pm} \bigl( g_2^\Charge \hat J_{r,+}^{0\dag}\hat J_{r',-}^0 + g_2^\SpinZ {\hat {\bf S}}_{r,+}^\dag \cdot {\hat {\bf S}}_{r',-}\pdag\bigr)\Bigr)
\end{split} 
\end{equation}
the density- and spin interaction part. The coupling contants are  
\begin{gather}
\label{Nodal coupling constants}
\begin{aligned}
&g_0^\Charge = a^2 \left( U/2 + 2V \right),& \quad & g_n^\Pair  = a^2 \left(U + 2V\left(1 + \cos \left(2Q\right)\right)\right) 
\\
&g_1^\Charge  = a^2 \left(U/2 + 2V\left(2 - \cos \left({2Q}\right) \right) \right),& \quad & g_2^\Charge  = a^2 \left( U/2 + V\left( 3 - \cos \left( {2Q} \right)\right) \right)
\\ 
&g_1^\SpinZ  =  - a^2 \left(2U + 8V\cos \left(2Q\right) \right),& \quad &g_2^\SpinZ  = - a^2 \left(2U + 4V\left( 1 + \cos \left( 2Q \right)\right) \right)
\end{aligned}
\end{gather} 
The antinodal part of \eqref{Effective Hamiltonian near half-filling} is given by 
\begin{equation}
\label{Regularised antinodal Hamiltonian}
\begin{split}
\Ha = &\sum\limits_{\alpha=\pm}\sum\limits_{r=\pm}\sum\limits_{\vk\in\Lambda_0^*} \tPiL^2 \left(\varepsilon_{r,0}(\vk) - \mu_0\right):\!\hat\psi^\dag_{r,0,\alpha}(\vk)\hat\psi\pdag_{r,0,\alpha}(\vk)\!: 
\\
&+\sum\limits_{r=\pm}\sum\limits_{\vp\in\tilde\Lambda^*}\frac{1}{L^2}  
\Bigl(g_a^\Charge \hat J_{r,0}^{0\dag}\hat J_{r,0}^0 + \tilde g_a^\Charge \hat J_{r,0}^{0\dag}\hat J_{-r,0}^0 + g_a^\SpinZ {\hat {\bf S}}_{r,0}^\dag \cdot {\hat {\bf S}}_{-r,0}\pdag + g_a^\Pair \hat \Pair_{r,0}^\dag \cdot \hat \Pair_{-r,0}\pdag\Bigr)
\end{split}
\end{equation}
with
\begin{equation}
\label{Effective antinodal chemical potential}
\mu_0 = \mu - 4t' -{U}/{2} + \left({U}/{4}+ V\right)\kappa^2 - \left({U}/{2}+ 4V\right)\nu
\end{equation}
the effective antinodal chemical potential, and
\begin{gather}
\label{Antinodal coupling constants}
\begin{aligned}
&g_a^\Charge  = a^2 \left(U/2 + 2V \right),& \quad & \tilde g_a^\Charge  = a^2\left(U/4 + 3V\right) 
\\
&g_a^\SpinZ  =  - a^2 \left( U - 4V \right),& \quad & g_a^\Pair  = a^2 \left( U - 4V \right)
\end{aligned}
\end{gather} 
the coupling constants. Note that we can write $\mu_0 \define \mu_{r,0}$ since the right hand side of \eqref{Effective antinodal chemical potential} is independent of $r$, and similarly $\Lambda_{0}^*\define\Lambda_{r,0}^*$.

Finally, the nodal fermions couple to the antinodal fermions through the following contribution to the effective Hamiltonian in \eqref{Effective Hamiltonian near half-filling} (note the abuse of duplicate notation in \eqref{Nodal-antinodal interaction: main} and \eqref{Regularised nodal-antinodal interaction})
\begin{equation}
\label{Regularised nodal-antinodal interaction}
\Hantnod = \HWZWantnod  + \frac{g_{na}^\Pair}{2}\sum\limits_{r,r',s=\pm}\sum\limits_{\vp\in\tilde\Lambda^*}\frac{1}{L^2}\chi_s\pdag(\vp) \Bigl(\hat\Pair_{r,s}^\dag \cdot \hat \Pair_{r',0}\pdag + \hat\Pair_{r',0}^\dag \cdot \hat \Pair_{r,s}\pdag\Bigr)
\end{equation}
with 
\begin{equation}
\label{Density- and spin part of nodal-antinodal interaction}
\HWZWantnod = \sum\limits_{r,r',s=\pm}\sum\limits_{\vp\in\tilde\Lambda^*}\frac{1}{L^2}\chi_s\pdag(\vp) \Bigl(g_{na}^\Charge \hat J_{r,s}^{0\dag} \hat J_{r',0}^0 + g_{na}^\SpinZ {\hat {\bf S}}_{r,s}^\dag\cdot {\hat {\bf S}}_{r',0}\pdag\Bigr)
\end{equation}
the density- and spin interaction part, and
\begin{gather}
\label{Nodal-antinodal coupling constants}
\begin{aligned}
&g_{na}^\Charge  = a^2 \left(U/2 + 4V \right),& \quad & g_{na}^\SpinZ  =  - 2a^2 U,& \quad & g_{na}^\Pair  = 2a^2 U
\end{aligned}
\end{gather}
the coupling constants.

\section{\label{Appendix Bosonization}Bosonization of nodal fermions -- additional details}

We collect without proofs some known results on non-abelian bosonization (Appendix~\ref{App:Non-abelian bosonization}); the reader is referred to Chapter 15 in \cite{DiFrancescoMathieuSenechal:1997} and references therein for further discussion. The notation used here is the same as that in Appendix~A of \cite{deWoulLangmann2:2010}. We also give the precise results on the bosonization of the nodal fermions (Appendices~\ref{App:Bosonization identities of nodal fermions}--\ref{Appendix: Bosonization of the nodal Hamiltonian}).

\subsection{\label{App:Non-abelian bosonization}Non-abelian bosonization}

Let $r,r'=\pm$ be chirality indices, $A,A'\in\mathcal{I}$ flavor indices with $\mathcal{I}$ some index set to be specified later, $\alpha,\alpha'=\pm$ spin indices, and $k,k'\in(2\pi/L)(\Z+1/2)$ 1D Fourier modes. We consider fermion operators $c^{(\dag)}_{r,A,\sigma}(k)$ defined on a fermion Fock space $\cF$ with normalized vacuum state $\Omega$ (Dirac sea) such that
\begin{equation}
\label{car} 
\{ c\pdag_{r,A,\alpha}(k),c^\dag_{r',A',\alpha'}(k')\} = \delta_{r,r'}\delta_{\alpha,\alpha'}\delta_{A,A'}\delta_{k,k'},\quad \{ c\pdag_{r,A,\alpha}(k),c\pdag_{r',A',\alpha'}(k')\}=0  
\end{equation}
and
\begin{equation}
\label{highest weight condition} 
c\pdag_{r,A,\alpha}(k)\Omega = c^\dag_{r,A,\alpha}(-k)\Omega  =0\quad \mbox{ for all }\; k\;\mbox{ such that }\; rk>0. 
\end{equation}
For $p\in(2\pi/L)\Z$ and $\mu=0,1,2,3$, let
\begin{equation}
\label{hatj}
\hat\jmath_{r,A}^{\mu}(p)\define \sum_{\alpha,\alpha'}\sum_{k\in\frac{2\pi}{L}(\Z+\frac12)}
:\! c^\dag_{r,A,\alpha}(k-p)\sigma_{\alpha,\alpha'}^\mu c\pdag_{r,A,\alpha'}(k)\!:
\end{equation}
with the colons denoting fermion normal ordering. These are well-defined operators on $\cF$ satisfying the commutation relations 
\begin{equation}
\label{Kac-Moody algebra}
\begin{split}
&\left[\hat\jmath_{r,A}^0(p),\hat\jmath_{r',A'}^0(p')\right]=2\delta_{r,r'}\delta_{A,A'}r\frac{Lp}{2\pi}\delta_{p+p',0}
\\
&\left[\hat\jmath_{r,A}^0(p),\hat\jmath_{r',A'}^i(p')\right]=0
\\
&\left[\hat\jmath_{r,A}^i(p),\hat\jmath_{r',A'}^j(p')\right]=2\delta_{r,r'}\delta_{A,A'}\Bigl(\sum_{k=1}^3\ii \epsilon_{ijk}\pdag\hat\jmath_{r,A}^k(p+p') + r\delta_{i,j}\frac{Lp}{2\pi}\delta_{p+p',0} \Bigr)
\end{split}
\end{equation} 
and 
\begin{equation}
\label{Current operator highest weight condition}
\hat\jmath_{r,A}^\mu(p)^\dag=\hat\jmath_{r,A}^\mu(-p), \qquad \hat\jmath_{r,A}^\mu(p)\Omega=0 \quad \mbox{ for all }\; p\;\mbox{ such that }\; rp\geq 0 .
\end{equation}  
Define also (generators of the Virasoro algebra \cite{DiFrancescoMathieuSenechal:1997})
\begin{equation}
\label{Virasoro generators}
\hat L_{r,A}\pdag(p)\define \sum_{\alpha}\sum_{k\in\frac{2\pi}{L}(\Z+\frac12)}
r(k-p/2):\! c^\dag_{r,A,\alpha}(k-p)c\pdag_{r,A,\alpha}(k)\!: \qquad (p\in(2\pi/L)\Z)
\end{equation}
such that
\begin{equation}
\label{Dirac Hamiltonian}
\sum_{A\in\mathcal{I}}\left[\hat L_{+,A}\pdag(0) + \hat L_{-,A}\pdag(0)\right] = \sum_{A\in\mathcal{I}}\sum_{r=\pm}\sum_{k\in\frac{2\pi}{L}(\Z+\frac12)} rk:\! c^\dag_{r,A,\alpha}(k)c\pdag_{r,A,\alpha}(k)\!:
\end{equation}
is proportional to an ordinary 1D (massless) Dirac Hamiltonian. The operators in \eqref{Virasoro generators} satisfy the commutation relations
\begin{equation}
\begin{split}
&\left[\hat L_{r,A}\pdag(p),\hat L_{r,A}\pdag(p')\right]=r (p-p')\hat L_{r,A}\pdag(p+p') + 2\frac{2\pi}{L}\delta_{p+p',0}\frac{1}{12}rp \Bigr[\Bigr(\frac{Lp}{2\pi}\Bigl)^2-1\Bigl]
\\
&\left[\hat L_{r,A}\pdag(p),\hat\jmath_{r,A}^\mu(p')\right]=-rp'\hat\jmath_{r,A}^\mu(p+p')
\end{split}
\end{equation} 
and $\hat L_{r,A}\pdag(p)\Omega=0$ if $rp\geq 0$. 
The following operator identity holds true (the Sugawara construction)
\begin{equation}
\label{Sugawara construction in 1D}
\hat L_{r,A}\pdag(p) =\frac{1}{4}\sum\limits_{p'\in\frac{2\pi}{L}\Z}\frac{2\pi}{L} \xxa \Bigl[\hat\jmath_{r,A}^0(p-p')\hat\jmath_{r,A}^0(p') + \frac{1}{3}\sum\limits_{i = 1}^3 \hat\jmath_{r,A}^i(p-p')\hat\jmath_{r,A}^i(p')\Bigr]\xxe
\end{equation}
with $\xxe\cdot\xxa$ denoting boson normal ordering as in \eqref{Boson normal-ordering}. 

\subsection{\label{App:Bosonization identities of nodal fermions}Bosonization identites for the nodal fermions}

The unspecified flavor index set $\mathcal{I}$ in Appendix~\ref{App:Non-abelian bosonization} is now defined as
\begin{equation}
\mathcal{I}\define \{ (s,x)\,:\, s=\pm,\, x\in\Lambda_{\mathrm{1D}} \}.
\end{equation} 
We can then represent the nodal fermion operators as
\begin{equation}
\label{psi from c} 
\hat\psi_{r,s,\alpha}(\vk)=\frac{L}{2\pi}\sqrt{\frac{\ta}{L}}\sum_{x\in \Lambda_{\mathrm{1D}}} c_{r,s,x,\alpha}(k_s)\ee^{-\ii k_{-s} x} \qquad (\vk=k_+\ve_{+}+k_{-}\ve_{-})
\end{equation} 
such that \eqref{car} and \eqref{highest weight condition} are equivalent to \eqref{Fermion flavor anticommutation relations: main} and \eqref{Highest Weight Condition: main}. 

\begin{proposition}
\label{Proposition spin-density operators:non-abelian}
The operators in \eqref{Nodal density- and spin operators} are well-defined operators on the fermion Fock space obeying the commutation relations ($\vp\in\tilde\Lambda^*$)
\begin{equation}
\label{Current algebra}
\begin{split}
\left[\hat J_{r,s}(\vp),\hat J_{r,s}(\vp')\right] & =  r\frac{4\pi p_s}{\ta}\Bigl(\frac{L}{2\pi}\Bigr)^2 \sum\limits_{n\in\Z}\delta_{\vp + \vp',2\pi n \ve_{-s}/\ta }
\\
\left[\hat S_{r,s}^i(\vp),\hat S_{r,s}^j(\vp')\right] & = \ii\sum\limits_{k=1}^3 \epsilon_{ijk}\hat S_{r,s}^k(\vp + \vp') 
+ \delta_{i,j} r\frac{\pi p_s}{\ta}\Bigl(\frac{L}{2\pi}\Bigr)^2 \sum\limits_{n\in\Z}\delta_{\vp + \vp',2\pi n \ve_{-s}/\ta }
\end{split}
\end{equation}
with all other commutators vanishing. Moreover, 
\begin{equation}
\hat J_{r,s}(\vp)^\dag = \hat J_{r,s}(-\vp), \quad \hat S_{r,s}^i(\vp)^\dag = \hat S_{r,s}^i(-\vp)
\end{equation}
and
\begin{equation}
\label{Boson highest-weight condition}
\hat J_{r,s}(\vp)\Omega = 0, \quad \hat S_{r,s}^i(\vp)\Omega = 0, \qquad \forall \vp\in\tilde\Lambda^*\, \mbox{ such that }\, rp_s\geq 0. 
\end{equation}
\end{proposition}
\begin{proof}
Using \eqref{psi from c} we can write the nodal density- and spin operators in terms of the operators in \eqref{hatj} as
\begin{equation}
\label{Nodal density- and spin operators from j} 
\begin{split}
\hat J_{r,s}(\vp) & = \sum_{x\in \Lambda_{\mathrm{1D}}}\hat\jmath_{r,s,x}^0(p_s)\ee^{-\ii p_{-s} x}
\\
\hat S_{r,s}^i(\vp) & =\frac{1}{2}\sum_{x\in \Lambda_{\mathrm{1D}}}\hat\jmath_{r,s,x}^i(p_s)\ee^{-\ii p_{-s} x}
\end{split} \qquad (\vp=p_+\ve_{+}+p_{-}\ve_{-}).
\end{equation}
The results stated in the proposition now follow by applying Equations~\eqref{Kac-Moody algebra}--\eqref{Current operator highest weight condition}.
\end{proof}

We define zero mode operators by
\begin{equation}
\label{hatNrs}
\hat{N}_{r,s,\alpha}(p_{-s})\define\sqrt{\frac{\ta}{2\pi}} \left. \hat{J}_{r,s,\alpha}(\vp)\right|_{p_s=0} \qquad (p_{-s}\in\tilde\Lambda_{1\rm D}^*)
\end{equation}
and their Fourier-transform
\begin{equation}
\label{Zero-mode operator}
N_{r,s,\alpha}(x) \define \sqrt{2\pi\ta} \sum_{p\in\tilde\Lambda^*_{1\mathrm{D}}}\frac{1}{L}   \hat{N}_{r,s,\alpha}(p)\ee^{\ii px} \qquad (x\in\Lambda_{1\rm D}).
\end{equation} 
When rewriting the nodal Hamiltonian in bosonized form in the next section, the following linear combinations of zero mode operators will be useful
\begin{equation}
\begin{split}
&Q_{\Charge;r,s}(x) \define \frac{1}{\sqrt 2}\sum\limits_{\alpha = \pm} \bigl[N_{+,s,\alpha}(x)+ rN_{-,s,\alpha}(x) \bigr] 
\\
&Q_{\SpinZ;r,s}(x) \define \frac{1}{\sqrt 2}\sum\limits_{\alpha = \pm} \alpha \bigl[N_{+,s,\alpha}(x)+ rN_{-,s,\alpha}(x) \bigr]
\end{split} \qquad (x\in\Lambda_{1\rm D}).
\end{equation}
We also define ${\hat Q}_{\Charge;r,s}(p)$ and ${\hat Q}_{\SpinZ;r,s}(p)$, $p\in\tilde\Lambda_{1\rm D}^*$, in a similar way (replace $N_{r,s,\alpha}(x)$ with ${\hat N}_{r,s,\alpha}(p)$ on the right hand sides above).

\begin{lemma}
\label{Lemma Fock space basis}
{\bf (a)} There exist unitary operators $R_{r,s,\alpha}(x)$ on the fermion Fock space commuting with all boson operators in \eqref{Boson operators from nodal density operators} and satisfying the commutation relations
\begin{equation}
\label{Klein factor commutator relations} 
\begin{split}
[N_{r,s,\alpha}(x),R_{r',s',\alpha'}(x')] &= r\delta_{r,r'}\delta_{s,s'}\delta_{\alpha,\alpha'}\delta_{x,x'}R_{r,s,\alpha}(x),\\
\{ R_{r,s,\alpha}(x),R_{r',s',\alpha'}(x')^\dag \} &= 2\delta_{r,r'}\delta_{s,s'}\delta_{\alpha,\alpha'}\delta_{x,x'} .  
\end{split}
\end{equation} 
\noindent {\bf (b)} Let $\cQ$ be the set of all pairs $(\vn,\vnu)$ with 
\begin{equation}
\vn=\left\{ n_{s,\alpha}(\vp) \right\}_{s,\alpha=\pm,\, \vp\in \hat\Lambda_s^*},\qquad 
\vnu=\left\{ \nu_{r,s,\alpha}(x) \right\}_{r,s,\alpha=\pm, \, x\in\Lambda_{1\mathrm{D}}}
\end{equation} 
and integers $\nu_{r,s,\alpha}(x)$ and $n_{s,\alpha}(\vp)\geq 0$ such that
\begin{equation}
\label{Finite condition}
\sum_{\alpha=\pm}\sum_{r,s=\pm}\sum_{x\in\Lambda_{1\mathrm{D}}}\nu_{r,s,\alpha}(x)^2 <\infty,\qquad \sum_{\alpha=\pm}\sum_{s=\pm}\sum_{\vp\in\hat\Lambda_s^*}|p_s| n_{s,\alpha}(\vp)<\infty .
\end{equation} 
Then the states
\begin{equation}
\label{Boson basis}
\eta_{\vn,\vnu}\define \Bigl( \prod_{\alpha=\pm}\prod_{s=\pm}\prod_{\vp\in\hat\Lambda_s^*} \frac{b_{s,\alpha}^\dag(\vp)^{n_{s,\alpha}(\vp)}}{\sqrt{n_{s,\alpha}(\vp)!}}\Bigr) \Bigl( \prod_{\alpha=\pm}\prod_{r,s=\pm}\prod_{x\in\Lambda_{1\mathrm{D}}} R_{r,s,\alpha}(x)^{\nu_{r,s,\alpha}(x)} \Bigr)\Omega,
\end{equation} 
with $(\vn,\vnu)\in\cQ$, provide a complete orthonormal basis in the fermion Fock space.\\
\noindent {\bf (c)} The states $\eta_{\vn,\vnu}$ are common eigenstates of the operators $N_{r,s,\alpha}(x)$ and $b_{s,\alpha}^\dag(\vp)b\pdag_{s,\alpha}(\vp)$ with eigenvalues $\nu_{r,s,\alpha}(x)$ and $n_{s,\alpha}(\vp)$, respectively.
\end{lemma} 
\noindent (\textit{Proof:} See the proof of Lemma~2.1 in \cite{deWoulLangmann2:2010}.) 

\begin{proposition}
\label{Proposition Fermions from Bosons}
For $r,s=\pm$, $\alpha=\pm$, $\vx\in\Lambda_s$, and $\epsilon>0$, the operator 
\begin{equation}
\label{psi and K 1}
\begin{split} 
\psi_{r,s,\alpha}\pdag(\vx;\epsilon)\define& \frac1{\sqrt{2\pi \ta\epsilon}}\ee^{\ii r\pi x_sN_{r,s,\alpha}(x_{-s})/L} R_{r,s,\alpha}(x_{-s})^{-r} \ee^{\ii r\pi x_s N_{r,s,\alpha}(x_{-s})/L} \\ & \times\exp\Bigl(r \frac{\ta}{2\pi} \sum_{\vp\in\hat\Lambda_{s}}\tPiL^2 \frac1{p_s} \hat{J}_{r,s,\alpha}(\vp)\ee^{\ii \vp\cdot\vx}\ee^{-\epsilon |p_s|/2} \Bigr) 
\end{split}
\end{equation}
is such that $\sqrt{2\pi \ta\epsilon}\psi_{r,s,\alpha}\pdag(\vx;\epsilon)$ is a unitary operator on the fermion Fock space, and
\begin{equation}
\label{psi from K} 
\hat\psi_{r,s,\alpha}\pdag(\vk) =\lim_{\epsilon\to 0^+} \frac1{2\pi} \sum_{x_{-s}\in\Lambda_{1\rm D}}\ta\int\limits_{-L/2}^{L/2}\ud x_s \,  \psi_{r,s,\alpha}\pdag(\vx;\epsilon)\ee^{-\ii\vk\cdot\vx}.
\end{equation} 
\end{proposition} 
\noindent (\textit{Proof:} See the proof of Proposition~2.2 in \cite{deWoulLangmann2:2010}.) 

The operator in \eqref{psi and K 1} yields a regularized version of the operator-valued distribution defined by the Fourier transform in \eqref{Position space operators}. This regularization is useful when computing correlation functions involving nodal operators (see \cite{deWoulLangmann2:2010} for further discussion of this).

The following proposition is key in bosonizing the nodal part of the effective Hamiltonian:
\begin{proposition}
\label{Proposition: Sugawara construction}
The following operator identities hold true
\begin{equation}
\label{Sugawara construction}
\begin{split}
\sum\limits_{\alpha=\pm}\sum\limits_{\vk\in\Lambda_s^*} \Bigl({\frac{{2\pi }}{L}}\Bigr)^2 rk_s :\hat\psi_{r,s,\alpha }^\dag(\vk)\hat\psi_{r,s,\alpha}\pdag(\vk): & = \ta\pi \sum\limits_{\alpha=\pm}\sum_{\vp\in\tilde\Lambda_s^*} \frac{1}{L^2} \xxa \hat{J}_{r,s,\alpha}^\dag \hat{J}_{r,s,\alpha}\pdag\xxe
\\
& = \ta\pi\sum\limits_{\vp\in \tilde \Lambda_s^*} \frac{1}{L^2} \xxa\Bigl(\frac{1}{2}\hat J_{r,s}^\dag\hat J_{r,s}\pdag + \frac{2}{3}{\hat{\bf S}}_{r,s}^\dag \cdot {\hat{\bf S}}_{r,s}\pdag \Bigr)\xxe 
\end{split}
\end{equation}
with all three expressions defining self-adjoint operators on the fermion Fock space.%; $\hat J_{r,s}^\dag \hat J_{r,s}\pdag = \hat J_{r,s}\pdag(-\vp) \hat J_{r,s}\pdag(\vp)$, etc.
\end{proposition}
\begin{proof}
See the proof of Proposition~2.1 in \cite{deWoulLangmann2:2010} for the first equality. The second equality is obtained using \eqref{Virasoro generators} and \eqref{Sugawara construction in 1D} for the special case $p=0$, together with relations \eqref{psi from c} and \eqref{Nodal density- and spin operators from j}.
\end{proof}

\subsection{\label{Appendix: Bosonization of the nodal Hamiltonian}Bosonization of the nodal Hamiltonian}

We write out the bosonization of the nodal Hamiltonian in \eqref{Regularised nodal Hamiltonian: app}--\eqref{WZW Hamiltonian: Interaction term: app} obtained from the extended Hubbard model. Using Proposition~\ref{Proposition: Sugawara construction}, we find
\begin{align}
\label{Charge part of nodal Hamiltonian}
\begin{split}
\HWZW_\Charge\pdag = \frac{v_F\pi\ta }{2}\xxa\Big(&\sum\limits_{r,s=\pm}\sum\limits_{\vp\in\tilde\Lambda_s^*} \frac{1}{L^2}\left(\left(1+ \gamma_0^\Charge\chi_s(\vp) \right)\hat J_{r,s}^{0\dag} {\hat J}_{r,s}^0 + \gamma_1^\Charge\chi_s(\vp)\hat J_{r,s}^{0\dag} {\hat J}_{-r,s}^0 \right)  
\\ 
&+ \sum\limits_{\vp\in\tilde\Lambda^*} \frac{1}{L^2}\gamma_2^\Charge\chi_+(\vp)\chi_-(\vp)\sum\limits_{r,r'=\pm}\hat J_{r,+}^{0\dag}{\hat J}_{r',-}^0 \Bigr) \xxe
\end{split}
\\
\label{Spin part of nodal Hamiltonian}
\begin{split}
\HWZW_\Spin\pdag =2{v_F}\pi\ta \xxa\Big(& \sum\limits_{r,s=\pm}\sum\limits_{\vp\in\tilde\Lambda_s^*} \frac{1}{L^2} \left(\hat {\bf S}_{r,s}^\dag  \cdot {\hat {\bf S}}_{r,s}\pdag/3 + \gamma_1^\SpinZ{\chi_s}(\vp)\hat {\bf S}_{r,s}^\dag\cdot {\hat {\bf S}}_{-r,s}\pdag \right)   
\\ 
&+ \sum\limits_{\vp\in\tilde\Lambda^*}\frac{1}{L^2}\gamma_2^\SpinZ{\chi_+}(\vp){\chi_-}(\vp) \sum\limits_{r,r'=\pm}\hat {\bf S}_{r,+}^\dag\cdot{\hat {\bf S}}_{r',-}\pdag\Bigr) \xxe
\end{split}
\end{align}
and where the (dimensionless) coupling constants are defined as (see also \eqref{Nodal coupling constants})
\begin{equation}
\label{Dimensionless coupling constants}
\gamma_0^\Charge \define \frac{2g_0^\Charge}{v_F \pi \ta}, \quad \gamma_1^\Charge \define \frac{g_1^\Charge}{v_F \pi \ta}, \quad \gamma_2^\Charge \define \frac{2g_2^\Charge}{v_F \pi \ta}, \quad \gamma_1^\SpinZ \define \frac{g_1^S}{4v_F\pi \ta}, \quad \gamma_2^\SpinZ  \define \frac{g_2^S}{2v_F \pi \ta}.
\end{equation}
We assume these satisfy
\begin{equation}
\label{Constraints on dimensionless coupling constants}
\left|{\gamma_1^\Charge}\right| < \left|{1+\gamma _0^\Charge}\right|, \quad \left|{\gamma_2^\Charge} \right| < \left| {1 + \gamma_0^\Charge + \gamma_1^\Charge} \right|, \quad \left| {\gamma_1^\SpinZ} \right| < 1, \quad \left| {\gamma_2^\SpinZ} \right| < \left| {1 + \gamma_1^\SpinZ} \right|,
\end{equation}
which implies the constraint
\begin{equation}
\label{Constraints on Hubbard parameters}
\frac{\left(3U + 4V\left[ 1 + 2\cos \left(2Q\right)\right]\right)\left(1-\kappa\right)}{8\pi\sin(Q)\left[t + 2t'\cos (Q)\right]} < 1.
\end{equation}
As in Section~\ref{Sec: An exactly solvable model of 2D electrons}, we write
\begin{equation}
\label{Break manifest SU(2)}
\begin{split}
\HWZW = \HMattis +\frac{1}{2}\sum\limits_{\vp\in\tilde\Lambda^*} \frac{1}{L^2} \Bigl( &g_1^S \sum\limits_{s =\pm} \chi_s(\vp){\bigl({\hat S_{+,s}^+(-\vp)\hat S_{-,s}^-(\vp) + h.c.}\bigr)}
\\
&+ g_2^S \sum\limits_{r,r'=\pm} \chi_+(\vp)\chi_-(\vp){\bigl( {\hat S_{r,+}^+(-\vp)\hat S_{r',-}^-(\vp) + h.c.} \bigr)}\Bigr) 
\end{split}
\end{equation}
with
\begin{equation}
\label{Spinfull Mattis Hamiltonian}
\HMattis = \HWZW_\Charge\pdag + \HWZW_\SpinZ\pdag
\end{equation}
and
\begin{align}
\label{Charge part of spinfull Mattis Hamiltonian}
\begin{split}
\HWZW_\Charge\pdag = &\frac{v_F}{2}\sum_{s=\pm}\sum_{\vp\in\hat\Lambda^*_s}\tPiL^2 \xxa\Bigl(\bigl(1+(\gamma_0^\Charge-\gamma_1^\Charge)\chi(\vp)\bigr)\hat\Pi^\dag_{\Charge;s}\hat\Pi\pdag_{\Charge;s}
\\
&+ \bigl(1+(\gamma_0^\Charge+\gamma_1^\Charge)\chi(\vp)\bigr)p_s^2\hat\Phi^\dag_{\Charge;s}\hat\Phi\pdag_{\Charge;s}+\gamma_2^\Charge p_+p_-\chi(\vp) \hat\Phi^\dag_{\Charge;s}\hat\Phi\pdag_{\Charge;-s}\Bigr)\xxe + H_{\Charge;z.m}\pdag 
\end{split}
\\
\label{Spin part of spinfull Mattis Hamiltonian}
\begin{split}
\HWZW_\SpinZ \define\, &\frac{v_F}{2}\sum_{s=\pm}\sum_{\vp\in\hat\Lambda^*_s}\tPiL^2 \xxa\Bigl(\bigl(1-\gamma_1^\SpinZ\chi(\vp)\bigr)\hat\Pi^\dag_{\SpinZ;s}\hat\Pi\pdag_{\SpinZ;s} 
\\
&+ \bigl(1+\gamma_1^\SpinZ\chi(\vp)\bigr)p_s^2\hat\Phi^\dag_{\SpinZ;s}\hat\Phi\pdag_{\SpinZ;s}+\gamma_2^\SpinZ p_+p_-\chi(\vp) \hat\Phi^\dag_{\SpinZ;s}\hat\Phi\pdag_{\SpinZ;-s}\Bigr)\xxe
+ H_{\SpinZ;z.m.}\pdag 
\end{split}
\end{align}
and where ($X=\Charge,\SpinZ$ and $\gamma_0^\SpinZ \define 0$)
\begin{equation}
\label{Zero mode terms of spinfull Mattis Hamiltonian}
\begin{split}
H_{X;z.m}\pdag =\, &\frac{v_F}{2}\tPiL^2\Bigl[\sum\limits_{s=\pm}\sum\limits_{\vp\in\hat\Lambda_s^*}  \xxa \bigl(\hat \Xi_{X;s}^\dag \hat\Phi_{X;s}\pdag + \hat\Phi_{X;s}^\dag  \hat\Xi_{X;s} \bigr)\xxe 
\\
&+\frac12\sum\limits_{r,s=\pm}\sum\limits_{p\in\tilde\Lambda_{1D}^* }  \bigl(1 + \gamma_0^X + r\gamma_1^X \bigr)\hat Q_{X;r,s}^\dag\hat Q_{X;r,s}\pdag  
+ \gamma_2^X \hat Q_{X;+,+}\pdag(0)\hat Q_{X;+,-}\pdag(0)\Bigr]
\end{split}
\end{equation}
\begin{equation}
\hat \Xi_{X;s}\pdag (\vp) \define - \frac{1}{\sqrt 2}\gamma_2^X \ii p_s \chi(\vp) \hat Q_{X;+,-s}\pdag(p_s )\delta_{p_{-s},0}
\end{equation}
denote terms involving zero mode operators; we have used the cutoff function in \eqref{Simplified cutoff} for simplicity. 

\begin{theorem}
\label{Theorem: Diagonalisation of Spinfull Mattis Hamiltonian}
There exists a unitary operator $\cU$ diagonalizing the Hamiltonian in \eqref{Spinfull Mattis Hamiltonian} as follows:
\begin{equation}
\cU^\dag \HMattis\cU = \sum_{s=\pm}\sum_{\vp\in\hat\Lambda_s^*}\left(\omega_{\Charge;s}\pdag(\vp) b_{\Charge;s}^\dag(\vp)b\pdag_{\Charge;s}(\vp) + \omega_{\SpinZ;s}\pdag(\vp) b_{\SpinZ;s}^\dag(\vp)b\pdag_{\SpinZ;s}(\vp)\right)+ \tilde H_{\ZM}+\cE^{(0)}
\end{equation}
with 
\begin{equation}
\label{Charge dispersion relation}
\omega_{\Charge;\pm}\pdag(\vp)= \begin{cases} \tilde{v}_{F}^\Charge\sqrt{\frac12\Bigl( |\vp|^2 \pm \sqrt{|\vp|^4-A_\Charge\bigl(2p_+p_-\bigr)^2 }\; \Bigr)}& \mbox{ if }\; \gamma_2^\Charge\chi(\vp)p_+p_-\neq 0
 \\v_F\sqrt{\bigl(1+\gamma_0^\Charge\chi(\vp)\bigr)^2-\bigl(\gamma_1^\Charge\chi(\vp)\bigr)^2}|p_\pm|& \mbox{ if }\;\gamma_2^\Charge\chi(\vp)p_+p_-= 0  \end{cases}
\end{equation} 
\begin{equation}
\label{Charge AvF} 
A_\Charge\define 1-\bigl[{\gamma_2^\Charge}/({1+\gamma_0^\Charge + \gamma_1^\Charge})\bigr]^2  ,\qquad \tilde{v}_F^\Charge\define v_F \sqrt{\bigl(1+\gamma_0^\Charge\bigr)^2-(\gamma_1^\Charge\bigr)^2}
\end{equation} 
and
\begin{equation}
\label{Spin dispersion relation}
\omega_{\SpinZ;\pm}\pdag(\vp)= \begin{cases} \tilde{v}_{F}^\SpinZ\sqrt{\frac12\Bigl( |\vp|^2 \pm \sqrt{|\vp|^4-A_\SpinZ\bigl(2p_+p_-\bigr)^2 }\; \Bigr)}& \mbox{ if }\; \gamma_2^\SpinZ\chi(\vp)p_+p_-\neq 0
 \\v_F\sqrt{1-\bigl(\gamma_1^\SpinZ\chi(\vp)\bigr)^2}|p_\pm|& \mbox{ if }\;\gamma_2^\SpinZ\chi(\vp)p_+p_-= 0  \end{cases}
\end{equation} 
\begin{equation}
\label{Spin AvF} 
A_\SpinZ\define 1-\bigl[{\gamma_2^\SpinZ}/({1 + \gamma_1^\SpinZ})\bigr]^2  ,\qquad \tilde{v}_F^\SpinZ\define v_F \sqrt{1-(\gamma_1^\SpinZ\bigr)^2}
\end{equation} 
the boson dispersion relations,
\begin{equation}
\label{Zero modes in diagonalised spinfull Mattis Hamiltonian}
\begin{split}
&\tilde H_{\ZM} = \frac{v_F\pi}{2L}\Biggl(\sum\limits_{s}\sum\limits_{x} \Bigl[\left(1 + \gamma_0^\Charge + \gamma_1^\Charge \right)A_\Charge\pdag Q_{\Charge;+,s}\pdag(x)^2 + \left(1 + \gamma_0^\Charge - \gamma_1^\Charge\right)Q_{\Charge;-,s}\pdag(x)^2 \Bigr] 
\\
&+ \frac{\ta}{L}\sum\limits_{s} \Biggl[\frac{\left(\gamma_2^\Charge\right)^2}{1 + \gamma_0^\Charge + \gamma_1^\Charge}\Bigl(\sum\limits_{x}Q_{\Charge;+,s}\pdag(x)\Bigr)^2 + \gamma_2^\Charge\Bigl(\sum\limits_{x}{Q_{\Charge; + ,s}\pdag(x)} \Bigr)\Bigl( \sum\limits_{x} Q_{\Charge; + , - s}\pdag (x) \Bigr) \Biggr] 
\\
&+\sum\limits_{s}\sum\limits_{x} \Bigl[\left(1 + \gamma_1^\SpinZ \right)A_\SpinZ\pdag Q_{\SpinZ;+,s}\pdag(x)^2 + \left(1 - \gamma_1^\SpinZ\right)Q_{\SpinZ;-,s}\pdag(x)^2 \Bigr] 
\\
&+ \frac{\ta}{L}\sum\limits_{s} \Biggl[\frac{\left(\gamma_2^\SpinZ\right)^2}{1 + \gamma_1^\SpinZ}\Bigl(\sum\limits_{x}Q_{\SpinZ;+,s}\pdag(x)\Bigr)^2 + \gamma_2^\SpinZ\Bigl(\sum\limits_{x}{Q_{\SpinZ; + ,s}\pdag(x)} \Bigr)\Bigl( \sum\limits_{x} Q_{\SpinZ; + , - s}\pdag (x) \Bigr) \Biggr] \Biggr)
\end{split}
\end{equation}
the part involving only zero mode operators (the sums are over $s=\pm$ and $x\in\Lambda_{1D}$),
and
\begin{equation}
\cE^{(0)} = \frac12\sum_{s=\pm}\sum_{\vp\in\hat\Lambda_s^*} \bigl(\omega_{\Charge;s}\pdag(\vp)+\omega_{\SpinZ;s}\pdag(\vp)-2v_F |p_s|\bigr)
\end{equation}
the groundstate energy of $\HMattis$.
\end{theorem}
\noindent (\textit{Proof:} See the proof of Theorem 3.1 in \cite{deWoulLangmann2:2010}.) 

Note that \eqref{Constraints on dimensionless coupling constants} are necessary and sufficient constraints on the coupling constants in order for $\cE^{(0)}$ to be well-defined and finite. One finds that the constraints on $\gamma_i^\Charge$, $i=0,1,2$, are always satisfied, while those on $\gamma_i^\SpinZ$, $i=1,2$, are fulfilled if \eqref{Constraints on Hubbard parameters} holds. 

\section{\label{App: Functional integration of nodal bosons}Functional integration of nodal bosons}

We give the results for the induced antinodal action obtained from the effective model in Appendix~\ref{App:Derivation of the effective model}. We truncate the nodal Hamiltonian in \eqref{Regularised nodal Hamiltonian: app} by only keeping $\HMattis$ (cf. \eqref{Break manifest SU(2)}), and then perform a similar truncation in the nodal-antinodal interaction \eqref{Density- and spin part of nodal-antinodal interaction}; we write
\begin{equation}
\HWZWantnod = H_{na}^{'(0)} + \frac12\sum\limits_{r,r',s=\pm}\sum\limits_{\vp\in\tilde\Lambda^*}\frac{1}{L^2}\chi(\vp) g_{na}^\SpinZ\Bigl( {\hat {S}}_{r,s}^+(-\vp) {\hat {S}}_{r',0}^-(\vp) + h.c.\Bigr)
\end{equation}
(using the simplified cutoff in \eqref{Simplified cutoff}). From \eqref{hatPihatPhi}, we find  
\begin{equation}
\begin{split}
H_{na}^{'(0)} = \sqrt{\frac{2}{\pi\ta}}\sum\limits_{r,s=\pm} \sum\limits_{\vp\in\hat\Lambda_s^*} \frac{1}{L^2}2\pi \ii p_s \chi(\vp)\Bigl( g_{na}^\Charge \hat J_{r,0}^{0\dag}\hat\Phi_{\Charge;s}\pdag +
\frac{g_{na}^\SpinZ}{2} \hat S_{r,0}^{3\dag}\hat \Phi_{\SpinZ;s}\pdag \Bigr) + z.m. 
\end{split}.
\end{equation}
The induced action becomes after integrating out the nodal bosons
\begin{equation}
\label{Induced action from abelian treatment}
S_{ind}^{(0)} \define \sum\limits_{n\in\Z}\sum\limits_{r,r'=\pm}\sum\limits_{\vp}\frac{1}{L^2}\left(\hat v_\Charge\pdag(\omega_n,\vp)\hat J_{r,0}^\dag\hat J_{r',0}\pdag + \hat v_\SpinZ\pdag(\omega_n,\vp) (\hat S_{r,0}^3)^\dag \hat S_{r',0}^3\right)
\end{equation}
with the density-density interaction potential 
\begin{equation}
\label{Induced charge interaction potential}
\hat v_\Charge\pdag(\omega_n,\vp) = -\frac{\left(g_{na}^\Charge\right)^2}{2\pi\ta v_F}\sum\limits_{s=\pm}\frac{W_{\Charge;s}\pdag(\vp)}{\omega_n^2 + \omega_{\Charge;s}\pdag(\vp)^2} \chi(\vp)
\end{equation}
where
\begin{equation}
\label{W charge}
W_{\Charge;\pm}\pdag(\vp) \define v_F^2\left(1 + \gamma_0^\Charge - \gamma_1^\Charge\right)\Biggl(\left|\vp\right|^2 \pm\frac{\left(p_+^2 - p_-^2\right)^2 + \sqrt{1-A_\Charge}\left(2p_+p_-\right)^2}{\sqrt{\left|\vp\right|^4 - A_\Charge \left(2p_+p_-\right)^2}}\Biggr)
\end{equation}
(see also definitions \eqref{Charge dispersion relation}--\eqref{Charge AvF}). Likewise, the induced spin-spin interaction potential is
\begin{equation}
\label{Induced spinZ interaction potential}
\hat v_\SpinZ\pdag(\omega_n,\vp) = -\frac{\left(g_{na}^\SpinZ\right)^2}{8\pi\ta v_F}\sum\limits_{s=\pm}\frac{W_{\SpinZ;s}\pdag(\vp)}{\omega_n^2 + \omega_{\SpinZ;s}\pdag(\vp)^2}\chi(\vp)
\end{equation}
with (see \eqref{Spin dispersion relation}--\eqref{Spin AvF})
\begin{equation}
\label{W spinZ}
W_{\SpinZ;\pm}\pdag(\vp) \define v_F^2\left(1 - \gamma_1^\SpinZ\right)\Biggl(\left|\vp\right|^2 \pm\frac{\left(p_+^2 - p_-^2\right)^2 - \sqrt{1-A_\SpinZ}\left(2p_+p_-\right)^2}{\sqrt{\left|\vp\right|^4 - A_\SpinZ \left(2p_+p_-\right)^2}}\Biggr)
\end{equation}
(the sign discrepancy between the numerators of \eqref{W charge} and \eqref{W spinZ} is due to the fact that $\gamma_2^\Charge\geq0$ while $\gamma_2^\SpinZ\leq 0$).

We also give the result when treating the nodal spin operators $\hat S_{r,s}^{i}$ as mutually commuting (to lowest order in $\ta$). Let
\begin{gather}
\label{Density- and spin boson operators: non-abelian case}
\begin{aligned}
\hat\Phi_{i;s}(\vp)\define&\sqrt{\frac{\ta}{2\pi}}\frac1{\ii p_s}\Bigl(\hat{S}_{+,s}^i(\vp) + \hat{S}_{-,s}^i(\vp) \Bigr), & \quad  \hat\Pi_{i;s}(\vp)\define&\sqrt{\frac{\ta}{2\pi}}\Bigl(-\hat{S}_{+,s}^i(\vp) + \hat{S}_{-,s}^i(\vp) \Bigr)
\end{aligned} 
\end{gather}
with $i=1,2,3$, $s=\pm$, and $\vp\in\hat\Lambda^*_s$; we note that $\hat\Phi_{3;s}\equiv\hat\Phi_{\SpinZ;s}$ and $\hat\Pi_{3;s}\equiv\hat\Pi_{\SpinZ;s}$ (cf. \eqref{hatPihatPhi}). 
Similar to $\HWZW_\SpinZ\pdag$ in \eqref{Spin part of spinfull Mattis Hamiltonian}, we can express $\HWZW_\Spin\pdag$ in \eqref{Spin part of nodal Hamiltonian} in terms of these operators as
\begin{equation}
\begin{split}
\HWZW_\Spin\pdag =\, &\frac{v_F}{2}\sum\limits_{i=1}^3\sum_{s=\pm}\sum_{\vp\in\hat\Lambda^*_s}\tPiL^2 \xxa\Bigl(\bigl[1/3-\gamma_1^\SpinZ\chi(\vp)\bigr]\hat\Pi^\dag_{i;s}\hat\Pi\pdag_{i;s} 
\\
&+ \bigl[1/3+\gamma_1^\SpinZ\chi(\vp)\bigr]p_s^2\hat\Phi^\dag_{i;s}\hat\Phi\pdag_{i;s}+\gamma_2^\SpinZ p_+p_-\chi(\vp) \hat\Phi^\dag_{i;s}\hat\Phi\pdag_{i;-s}\Bigr)\xxe +z.m.
\end{split}
\end{equation}
Likewise, the density- and spin part of the nodal-antinodal interaction given in \eqref{Density- and spin part of nodal-antinodal interaction} can be written as
\begin{equation}
\HWZWantnod = \sqrt{\frac{2}{\pi\ta}}\sum\limits_{r,s=\pm} \sum\limits_{\vp\in\hat\Lambda_s^*} \frac{1}{L^2}2\pi \ii p_s \chi(\vp)\Bigl( g_{na}^\Charge \hat J_{r,0}^\dag\hat\Phi_{\Charge;s}\pdag + \frac{g_{na}^\SpinZ}{2}\sum\limits_{i = 1}^3 (\hat S_{r,0}^i)^\dag\hat \Phi_{i;s}\pdag \Bigr) + z.m. 
\end{equation} 
The induced action is then
\begin{equation}
S_{ind} \define \sum\limits_{n\in\Z}\sum\limits_{r,r'=\pm}\sum\limits_{\vp \in \tilde\Lambda^*}\frac{1}{L^2}\left(\hat v_\Charge\pdag(\omega_n,\vp)\hat J_{r,0}^\dag\hat J_{r',0}\pdag + \hat v_\Spin\pdag(\omega_n,\vp) \hat\Spin_{r,0}^\dag \cdot \hat\Spin_{r',0}\pdag\right) 
\end{equation}
where the spin-spin interaction potential is now given by
\begin{equation}
\label{Induced spin interaction potential}
\hat v_\Spin\pdag(\omega_n,\vp) = -\frac{\left(g_{na}^\SpinZ\right)^2}{8\pi\ta v_F}\sum\limits_{s=\pm}\frac{W_{\Spin;s}\pdag(\vp)}{\omega_n^2 + \omega_{\Spin;s}\pdag(\vp)^2}\chi(\vp)
\end{equation}
with
\begin{equation}
\label{Dispersion relation with spin rotation invariance}
\begin{split}
&\omega_{\Spin;\pm}\pdag(\vp) \define \tilde{v}_{F}^\Spin\sqrt{\frac12\Bigl( |\vp|^2 \pm \sqrt{|\vp|^4-A_\Spin\bigl(2p_+p_-\bigr)^2 }\; \Bigr)}
\\
&A_\Spin\define 1-\bigl[{\gamma_2^\SpinZ}/({1/3 + \gamma_1^\SpinZ})\bigr]^2  ,\qquad \tilde{v}_F^\Spin\define v_F \sqrt{(1/3)^2-(\gamma_1^\SpinZ\bigr)^2}
\end{split}
\end{equation}
(note that this differ from \eqref{Spin dispersion relation}--\eqref{Spin AvF}), and
\begin{equation}
\label{W}
W_{\Spin;\pm}\pdag(\vp) \define v_F^2\left(1/3 - \gamma_1^\SpinZ\right)\Biggl(\left|\vp\right|^2 \pm\frac{\left(p_+^2 - p_-^2\right)^2 - \sqrt{1-A_\Spin}\left(2p_+p_-\right)^2}{\sqrt{\left|\vp\right|^4 - A_\Spin \left(2p_+p_-\right)^2}}\Biggr).
\end{equation}
For \eqref{Induced spin interaction potential}{\em ff} to be well-defined, we need to impose the somewhat stricter conditions on the coupling constants
\begin{equation}
\left| {\gamma_1^\SpinZ} \right| < 1/3, \qquad \left| {\gamma_2^\SpinZ} \right| < \left| {1/3 + \gamma_1^\SpinZ} \right|
\end{equation}
which translates into (cf. \eqref{Constraints on Hubbard parameters})
\begin{equation}
\frac{\left(3U + 4V\left[ 1 + 2\cos \left(2Q\right)\right]\right)\left(1-\kappa\right)}{8\pi\sin(Q)\left[t + 2t'\cos (Q)\right]} < \frac{1}{3}.
\end{equation}

\end{document}